\RequirePackage[2020-02-02]{latexrelease}
\documentclass[journal,hidelinks]{IEEEtran}

\usepackage{amsmath,amssymb,amsfonts}
\usepackage{algorithmic}
\usepackage{graphicx}
\usepackage{textcomp}
\usepackage{caption}
\usepackage[normalem]{ulem}
\usepackage{subcaption}
\usepackage[inline]{enumitem}
\usepackage{amsmath}
\usepackage{amsfonts}

\usepackage{mathtools}
\usepackage{threeparttable}
\usepackage{times}
\usepackage{fancyhdr,graphicx,amsmath,amssymb}
\usepackage[ruled,vlined]{algorithm2e}
\include{pythonlisting}
\usepackage[nolist]{acronym} 
\usepackage{comment}
\usepackage{booktabs}
\usepackage{bm} 
\usepackage{multirow}
\usepackage[utf8]{inputenc}
\DeclareUnicodeCharacter{1EF3}{\`y}
\DeclarePairedDelimiter\ceil{\lceil}{\rceil}
\DeclarePairedDelimiter\floor{\lfloor}{\rfloor}
\usepackage[style=ieee, backend=biber, maxnames=6]{biblatex}
\AtEveryBibitem{%
  \clearlist{language}
}

\usepackage{lipsum}
\usepackage{fancyhdr}

\fancypagestyle{sec}{\lhead{\tmpx}}

\fancypagestyle{arXiv}
{
   \fancyhf{}
   
   \chead{\textcolor{gray}{This paper has been accepted for publication at the \\ IEEE Transactions on Neural Networks and Learning Systems,  2022}}
   
   \cfoot{\vspace{-6mm} \fontsize{=6pt}{10pt}\selectfont  \textcolor{gray}{© 2022 IEEE. Personal use of this material is permitted. Permission from IEEE must be obtained for all other uses, in any current or future media, including reprinting/republishing this material for advertising or promotional purposes, creating new collective works, for resale or redistribution to servers or lists, or reuse of any copyrighted component of this work in other works}\\\fontsize{=10pt}{12pt}\selectfont\thepage}
   
}

\usepackage[draft]{pdfcomment} 
\usepackage[hyperfirst=false,acronym,sort=none,shortcuts,nopostdot,style=super,nonumberlist,toc,nogroupskip]{glossaries}
\usepackage[usenames,dvipsnames]{xcolor} 
\glsdisablehyper 

\newcommand{\tipcolor}{Black} 

\newcommand{\accolor}[1]{\textcolor{\tipcolor}{#1}}

\newacronymstyle{myacro}
{%
  \GlsUseAcrEntryDispStyle{long-short}%
}%
{%
  \GlsUseAcrStyleDefs{long-short}%
}

\setacronymstyle{myacro}

\newcommand*{\tip}[1]{
    \ifglsused{#1}{
      {\pdftooltip{\accolor{\glsentryshort{#1}}}{\glsentrydesc{#1}}}%
    }{%
      \gls{#1}
    }%
}%

\newcommand*{\tipshort}[1]{
      {\pdftooltip{\accolor{\glsentryshort{#1}}}{\glsentrydesc{#1}}}%
}%

\newcommand*{\tips}[1]{
    \ifglsused{#1}{
      {\pdftooltip{\accolor{\glsentryshortpl{#1}}}{\glsentrydescplural{#1}}}%
    }{%
      \glspl{#1}
    }%
}%

\newacronym{aae}{AAE}{Average Angular Error}
\newacronym{abmof}{ABMOF}{Adaptive Block Matching Optical Flow}
\newacronym{accmem}{ACC Mem}{Accumulation Memory}
\newacronym{accreg}{ACC REG}{Accumulation Register}
\newacronym{acp}{ACP}{Accelerator Coherency Port}
\newacronym{adc}{ADC}{Analog-to-Digital Converter}
\newacronym{add}{ADD}{adder}
\newacronym{admm}{ADMM}{Alternating Direction Method of Multipliers}
\newacronym{aee}{AEE}{Average Endpoint Error}
\newacronym{aer}{AER}{Address-event-representation}
\newacronym{afe}{AFE}{Analog Front-End}
\newacronym{ai}{AI}{Artificial Intellegence}
\newacronym{alm}{ALM}{Adaptive Logic Module}
\newacronym{ampro}{AMPRO}{Advanced Mechanical Bipedal Experimental Robotics Prosthesis}
\newacronym{am}{AM}{acoustic model}
\newacronym{ann}{ANN}{Artificial Neural Network}
\newacronym{aps}{APS}{Active Pixel Sensor}
\newacronym{ap}{AP}{Activation Pipeline}
\newacronym{asic}{ASIC}{Application Specific Integrated Circuit}
\newacronym{at}{AT}{Adder Trees}
\newacronym{axis}{AXIS}{AXI-Stream}
\newacronym{axi}{AXI}{Advanced eXtensible Interface}
\newacronym{bbb}{BBB}{Beaglebone Black}
\newacronym{bbs}{BBS}{Bank Balanced Sparsity}
\newacronym{blen}{BLEN}{Burst Length}
\newacronym{bnn}{BNN}{Binary Neural Network}
\newacronym{bpf}{BPF}{Band-Pass Filter}
\newacronym{bptt}{BPTT}{Backpropagation Through Time}
\newacronym{bram}{BRAM}{Block Random Access Memory}
\newacronym{br}{BR}{Balance Ratio}
\newacronym{cbcsc}{CBCSC}{Column-Balanced Compressed Sparse Column}
\newacronym{cbtd}{CBTD}{Column-Balanced Targeted Dropout}
\newacronym{ccd}{CCD}{Charge-coupled Device}
\newacronym{cfg}{CFG}{configuration}
\newacronym{clk}{CLK}{clock}
\newacronym{cmos}{CMOS}{Complementary metal–oxide–semiconductor}
\newacronym{cnn}{CNN}{Convolutional Neural Network}
\newacronym{cnt}{CNT}{counter}
\newacronym{conv}{CONV}{convolutional}
\newacronym{cots}{COTS}{Commodity Off-The-Shelf}
\newacronym{cpld}{CPLD}{Complex Programmable Logic Device}
\newacronym{cpu}{CPU}{Central Processing Unit}
\newacronym{cp}{CP}{column pointer}
\newacronym{csc}{CSC}{Compressed Sparse Column}
\newacronym{csr}{CSR}{Compressed Sparse Row}
\newacronym{ctc}{CTC}{Connectionist Temporal Classification}
\newacronym{ctrl}{CTRL}{Controller}
\newacronym{cuda}{CUDA}{Compute Unified Device Architecture} 
\newacronym{cudnn}{cuDNN}{CUDA Deep Neural Network library}
\newacronym{cv}{CV}{Computer Vision}
\newacronym{dac}{DAC}{Digital-to-Analog Converter}
\newacronym{das}{DAS}{Dynamic Audio Sensor}
\newacronym{davis}{DAVIS}{Dynamic and Active pixel Vision Sensor}
\newacronym{dbe}{DBE}{Digital Back-End}
\newacronym{ddr}{DDR}{Double Data Rate}
\newacronym{dec}{DEC}{Decoder}
\newacronym{dev}{DEV}{development}
\newacronym{deltagru}{DeltaGRU}{Delta Gated Recurrent Unit}
\newacronym{deltalstm}{DeltaLSTM}{Delta Long-Short Term Memory}
\newacronym{dl}{DL}{Deep Learning}
\newacronym{dma}{DMA}{Direct Memory Access}
\newacronym{dm}{DM}{Delta Memory}
\newacronym{dnn}{DNN}{Deep Neural Network}
\newacronym{dn}{DN}{Delta Network}
\newacronym{dpe}{DPE}{Delta Processing Element}
\newacronym{dram}{DRAM}{Dynamic Random Access Memory}
\newacronym{dsp}{DSP}{Digital Signal Processing}
\newacronym{du}{DU}{Delta Unit}
\newacronym{dvs}{DVS}{Dynamic Vision Sensor}
\newacronym{edflow}{EDFLOW}{Event-driven Optical Flow}
\newacronym{eds}{EDS}{Event-Driven System}
\newacronym{eie}{EIE}{Efficient Inference Engine}
\newacronym{emmc}{eMMC}{Embedded Multi Media Card}
\newacronym{ese}{ESE}{Efficient Speech Recognition Engine}
\newacronym{fast}{FAST}{Features from Accelerated Segment Test}
\newacronym{fcl}{FCL}{Fully-Connected Layer}
\newacronym{fc}{FC}{Fully-Connected}
\newacronym{fex}{FEx}{Feature Extractor}
\newacronym{fft}{FFT}{Fast Fourier Transform}
\newacronym{fifo}{FIFO}{First-In First-Out}
\newacronym{fllatc}{FLL-ATC}{Frequency locked-loop analog-to-time converter}
\newacronym{fmllr}{fMLLR}{feature-space maximum likelihood linear regression}
\newacronym{fp16}{FP16}{16-bit floating-point number}
\newacronym{fp32}{FP32}{32-bit floating-point number}
\newacronym{fp8}{FP8}{8-bit floating-point number}
\newacronym{fpga}{FPGA}{Field Programmable Gate Array}
\newacronym{fps}{FPS}{Frame Per Second}
\newacronym{fp}{FP}{floating-point}
\newacronym{fsm}{FSM}{Finite-State Machine}
\newacronym{fxp}{FXP}{fixed-point}
\newacronym{gb}{GB}{gigabytes}
\newacronym{gemm}{GEMM}{General Matrix Multiplication}
\newacronym{gpio}{GPIO}{General-Purpose Input/Output}
\newacronym{gpu}{GPU}{Graphics Processing Unit}
\newacronym{gp}{GP}{General Purpose}
\newacronym{gru}{GRU}{Gated Recurrent Unit}
\newacronym{gscd}{GSCD}{Google Speech Command Dataset}
\newacronym{gt}{GT}{Ground Truth}
\newacronym{hbm}{HBM}{High Bandwidth Memory}
\newacronym{hcgs}{HCGS}{Hierarchical Coarse-Grain Sparsity}
\newacronym{hdl}{HDL}{Hardware Description Language}
\newacronym{hdr}{HDR}{high dynamic range}
\newacronym{hls}{HLS}{High Level Synthesis}
\newacronym{hmc}{HMC}{Hybrid Memory Cube}
\newacronym{hmm}{HMM}{Hidden Markov Model}
\newacronym{hpe}{HPE}{Heterogeneous Processing Element}
\newacronym{hp}{HP}{High Performance}
\newacronym{ht}{HT}{High Throughput}
\newacronym{idx}{IDX}{index}
\newacronym{ieu}{IEU}{Input Encoding Unit}
\newacronym{iir}{IIR}{Infinite Impulse Response}
\newacronym{imc}{IMC}{In-Memory Computing}
\newacronym{imu}{IMU}{Inertial Measurement Unit}
\newacronym{int16}{INT16}{16-bit integer}
\newacronym{int8}{INT8}{8-bit integer}
\newacronym{int}{INT}{integer}
\newacronym{iot}{IoT}{Internet-of-Things}
\newacronym{io}{I/O}{Input/Output}
\newacronym{ipu}{IPU}{Input Processing Unit}
\newacronym{ip}{IP}{Intellectual Property}
\newacronym{isa}{ISA}{Instruction Set Architecture}
\newacronym{jtag}{JTAG}{Joint Test Action Group}
\newacronym{kb}{KB}{kilobytes}
\newacronym{kws}{KWS}{Keyword Spotting}
\newacronym{lasso}{LASSO}{least absolute shrinkage and selection operator}
\newacronym{ldo}{LDO}{Low-dropout}
\newacronym{lidx}{LIDX}{local index}
\newacronym{ll}{LL}{Low Latency}
\newacronym{lstm}{LSTM}{Long Short-Term Memory}
\newacronym{lutram}{LUTRAM}{Look-up Table based Random Access Memory}
\newacronym{lut}{LUT}{look-up table}
\newacronym[description={Multiply-Accumulate is the basic operation of signal processing and artificial neural networks. One MAC is 2 Op.}]{mac}{MAC}{Multiply-Accumulate}
\newacronym{mb}{MB}{megabytes}
\newacronym{mcu}{MCU}{Microcontroller Unit}
\newacronym{mfcc}{MFCC}{Mel Frequency Cepstral Coefficients}
\newacronym{mkl}{MKL}{Math Kernel Library}
\newacronym{mlp}{NLP}{Multi-Layer Perception}
\newacronym{mm2s}{MM2S}{Memory Mapped to Stream}
\newacronym{mmp}{MMP}{Mini-Module-Plus}
\newacronym{mul}{MUL}{multiplier}
\newacronym{mxspv}{MxSPV}{Matrix-Sparse Vector Multiplication}
\newacronym{mxv}{MxV}{Matrix-Vector Multiplication}
\newacronym{nas}{NAS}{Neural Architecture Search}
\newacronym{ncs}{NCS}{Neural Compute Stick}
\newacronym{nlp}{NLP}{Natural Language Processing}
\newacronym{norm}{Norm}{normalization}
\newacronym{nss}{NSS}{Nyquist-Sample System}
\newacronym{nzil}{NZIL}{Non-Zero Index List}
\newacronym{nzi}{NZI}{Nonzero Index}
\newacronym{nzvl}{NZVL}{Non-Zero Value List}
\newacronym{nzv}{NZV}{Nonzero Value}
\newacronym{ob}{OBUF}{Output Buffer}
\newacronym{onn}{ONN}{Optical Neural Network}
\newacronym{os}{OS}{Operating System}
\newacronym{pcb}{PCB}{Printed Circuit Board}
\newacronym{pcol}{pcol}{weight column pointers}
\newacronym{pc}{PC}{Personal Computer}
\newacronym{pd}{PD}{proportional–derivative}
\newacronym{per}{PER}{Phone Error Rate}
\newacronym{pe}{PE}{Processing Element}
\newacronym{pfm}{PFM}{Pulse-frequency-modulated}
\newacronym{pl}{PL}{Programmable Logic}
\newacronym{prm}{PRM}{Pixel Rendering Module}
\newacronym{ps}{PS}{Processing System}
\newacronym{pt}{PT}{Previous Time}
\newacronym{pwm}{PWM}{Pulse-Width-Modulated}
\newacronym{qspi}{QSPI}{Quad Serial Peripheral Interface}
\newacronym{ralut}{RALUT}{Range Addressable Lookup Table}
\newacronym{ram}{RAM}{Random Access Memory}
\newacronym{ratp}{RATP}{Recursive Adaptive Temporal Pooling}
\newacronym{rec}{Rec}{Rectified}
\newacronym{relu}{ReLU}{Rectified Linear Unit}
\newacronym{reram}{ReRAM}{Resistive random-access memory}
\newacronym{ridx}{RIDX}{relative index}
\newacronym{risc}{RISC}{Reduced Instruction Set Computer}
\newacronym{rmse}{RMSE}{Root-Mean-Square Error}
\newacronym{rnn}{RNN}{Recurrent Neural Network}
\newacronym{ros}{ROS}{Robot Operating System}
\newacronym{s2mm}{S2MM}{Stream to Memory Mapped}
\newacronym{sad}{SAD}{Sum of Absolute Differences}
\newacronym{sae}{SAE}{Surface of Active Events}
\newacronym{sdk}{SDK}{Software Development kit}
\newacronym{sd}{SD}{Secure Digital}
\newacronym{se}{SE}{Single-Ended}
\newacronym{sits}{SITS}{Speed Invariant Time Surface}
\newacronym{smem}{SMEM}{State Memory}
\newacronym{sm}{SM}{Supplementary Material}
\newacronym{snn}{SNN}{Spiking Neural Network}
\newacronym{snr}{SNR}{Signal-to-Noise Ratio}
\newacronym{soc}{SoC}{System-on-a-Chip}
\newacronym{som}{SOM}{System-on-Module}
\newacronym{spi}{SPI}{Serial Peripheral Interface}
\newacronym{spmxspv}{SPMxSPV}{Sparse Matrix-Sparse Vector Multiplication}
\newacronym{spmxv}{SPMxV}{Sparse Matrix-Vector Multiplication}
\newacronym{sram}{SRAM}{Static Random Access Memory}
\newacronym{sro}{SRO}{Switched-Ring-Oscillator}
\newacronym{test}{TEST}{test}
\newacronym{tf}{TF}{Threshold Function}
\newacronym{tsmc}{TSMC}{Taiwan Semiconductor Manufacturing Company}
\newacronym{usb}{USB}{Universal Serial Bus}
\newacronym{vad}{VAD}{Voice Activity Detection}
\newacronym{val}{VAL}{Value}
\newacronym{wer}{WER}{Word Error Rate}
\newacronym{wfst}{WFST}{Weighted Finite State Transducer}
\newacronym{wl}{WL}{workload}
\newacronym{wmem}{WMEM}{Weight Memory}
\newacronym{xor}{XOR}{Exclusive-OR}

\begin{document}

\title{Spartus: A 9.4 TOp/s FPGA-based LSTM Accelerator Exploiting Spatio-Temporal Sparsity}

\author{Chang~Gao,~\IEEEmembership{Member,~IEEE,}
        Tobi~Delbruck,~\IEEEmembership{Fellow,~IEEE}
        and~Shih-Chii~Liu,~\IEEEmembership{Fellow,~IEEE}
\thanks{C. Gao, T. Delbruck and S-C. Liu are with the Sensors Group at the Institute of Neuroinformatics, University of Zurich and ETH Zurich, Zurich (\href{https://sensors.ini.uzh.ch}{sensors.ini.uzh.ch}), Switzerland, 8057 Switzerland (email: 	
chang@ini.uzh.ch; shih@ini.uzh.ch; tobi@ini.uzh.ch).}
\thanks{This work was supported by Samsung Advanced Institute 
of Technology (SAIT) Global Research \textit{Neuromorphic Processor} project, the Swiss National Competence Center in Robotics (NCCR Robotics, 51NF40\_185543), and the Swiss National Science Foundation BRIDGE project VIPS (181010). 
}
}


\maketitle

\begin{abstract}
Long Short-Term Memory (LSTM) recurrent networks are frequently used for tasks involving time-sequential data such as speech recognition. Unlike previous LSTM accelerators that either exploit spatial weight sparsity or temporal activation sparsity,  this paper proposes a new accelerator called ``Spartus" that exploits spatio-temporal sparsity to achieve ultra-low latency inference.
Spatial sparsity is induced using a new Column-Balanced Targeted Dropout (CBTD) structured pruning method, producing structured sparse weight matrices for a balanced workload. 
The pruned networks running on Spartus hardware achieve weight sparsity levels of up to 96\% and 94\% with negligible accuracy loss on the TIMIT and the Librispeech datasets.
To induce temporal sparsity in LSTM, we  
extend the previous DeltaGRU method to the DeltaLSTM method.
Combining spatio-temporal sparsity with CBTD and DeltaLSTM saves on weight memory access and associated arithmetic operations.
The Spartus architecture is scalable and supports real-time online speech recognition when implemented on small and large FPGAs.
Spartus per-sample latency for a single DeltaLSTM layer of 1024 neurons averages 1 \textmu s. 
Exploiting spatio-temporal sparsity on our test LSTM network using the TIMIT dataset leads to 46$\times$ speedup of Spartus over its theoretical hardware performance to achieve 9.4 TOp/s effective batch-1 throughput and 1.1 TOp/s/W power efficiency.
\end{abstract}


%
\IEEEpeerreviewmaketitle

\section{Introduction}
\thispagestyle{arXiv}
\IEEEPARstart{R}{ecurrent} Neural Networks (\textbf{\tipshort{rnn}}s) are widely used in tasks that involve temporal sequences. \tipshort{rnn} variants such as the \tip{lstm}~\cite{lstm_hoch97} and \tip{gru}~\cite{Cho2014} models use additional gating units to mitigate the problem of vanishing gradient. These variants achieve state-of-the-art prediction accuracy in tasks involving input temporal  sequences such as automatic speech recognition~\cite{Amodei2015,Graves2013}, and natural language processing~\cite{mikolov2010}. 
RNNs are also useful in latency-critical real-time control tasks, such as robotic prosthesis control~\cite{GaoICRA2020}, gaming AI~\cite{Vinyals2019} and autonomous driving~\cite{Mozaffari2020}, which require the hardware to maintain high throughput even with a batch size of one input sample.

Batch-1 \tipshort{rnn} calculation is dominated by \tip{mxv} since there is only one input stream and the fetched big weight matrix cannot be shared across multiple input streams.
The computational cost of this calculation grows quadratically with a linear increase in \tipshort{rnn} units.
The reasons why it is difficult to achieve low cost, latency, and power simultaneously for \tipshort{rnn} hardware inference are fourfold. 
First, the temporal dependence between the current and previous network output creates a critical path that limits the parallelism between time steps. 
Second, large networks are essential for high accuracy, leading to a large memory footprint that is expensive to buffer on-chip. 
Third, \tip{mxv} is a memory-bounded operation, and the available memory bandwidth limits the minimum latency to fetch the large weight matrices.
Fourth, memory access consumes at least 10$\times$ more energy than arithmetic operations with the same number of bits~\cite{Horowitz2014,Jouppi2021}. 
In short, the key to achieving low-cost, low-latency, low-power \tipshort{rnn} inference is to reduce the memory bottleneck, that is, to minimize the needed access of weights.
\begin{figure}[!t]
	\centering
	\includegraphics[width=0.95\linewidth]{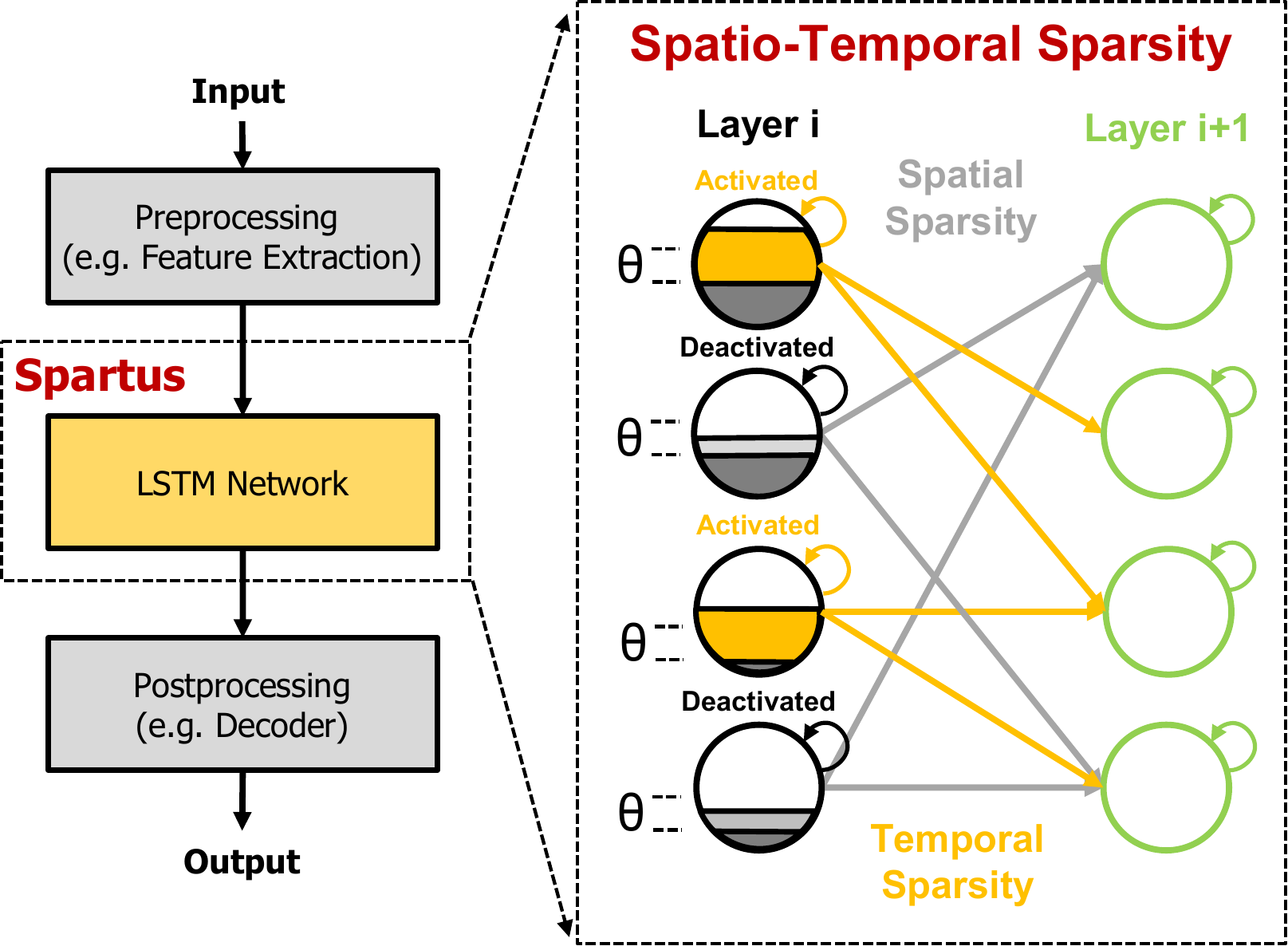}
	\caption{Concept of the Spartus accelerator that exploits spatio-temporal sparsity on layer $i$ of a multi-layer \tip{lstm} network. The structured spatial sparsity ensures an equal number of outward connections between units in layer $i$ for a balanced workload. The change of activations (yellow) can be either positive or negative.}
	\label{Fig:concept}
\end{figure}

A popular method to reduce memory access is to sparsify the weight matrices.  
Pruning methods~\cite{Han2015,han2017ese,Gomez2018} remove unimportant connections between neurons, resulting in sparse weight matrices with a smaller memory footprint than the original dense matrices. 
Structured pruning was later introduced in \tipshort{rnn} accelerators to address the problem of workload imbalance caused by irregular weight sparsity patterns after pruning~\cite{wen2018learning, Cao2019, Wang2019,Kadetotad2020}. 
The \tip{bbs}~\cite{Cao2019} method divides the rows of a weight matrix into banks of equal length. 
Then, fine-grain pruning is applied to each bank with an equal number of nonzero elements in each bank.
The \tip{hcgs} structured pruning method pre-partitions the weight matrix with hierarchical levels of squared submatrices and removes equal numbers of weight elements in each submatrix to ensure a balanced workload. This method was used in a recent power-efficient \tip{asic} \tip{rnn} accelerator~\cite{Kadetotad2020}.
Structured weight matrices can also reduce the effective number of weights  and \tip{mxv} cost without increasing sparsity: 
Block-circulant matrices enable the use of the fast Fourier transform to reduce the cost of \tip{mxv} from $\mathcal{O}(n^2)$ to $\mathcal{O}(n\log(n))$~\cite{Wang2017,Wang2018, Li2019}. 

Another way to reduce memory access requirements is to increase the temporal sparsity of activations, as in spiking neural networks. The \tip{dn} algorithm was inspired by the neuromorphic principle that neurons have sparse activity transmission and introduced as a method to induce temporal sparsity in deep networks by replacing state vectors with delta vectors that contain the deltas of the temporal difference of the states between two adjacent time steps~\cite{neil2016delta}. Zeroing deltas that are below a delta threshold produces sparse delta vectors. Using these sparsified delta vectors results in an insignificant accuracy loss if the \tipshort{rnn} is properly trained. Using 
hardware that can skip zeros in delta state vectors, we can remove operations from entire columns of the weight matrix, which is intrinsically workload balanced~\cite{GaoDeltaRNN2018,edgedrnn}. 
Temporal sparsity is similar to activation sparsity, and to our best knowledge, it was only exploited in 3 previous hardware accelerator works on \tips{cnn}~\cite{aimar2019,chen2020} and \tips{rnn}~\cite{GaoDeltaRNN2018}.

Finally, quantization of weights and states~\cite{xu2018alternating} can be combined with the previous sparsity-inducing methods to reduce the model size further. 
In the basic \tip{dn} algorithm, columns of weights that are not skipped are still dense. 
There is an opportunity to further reduce the memory footprint by inducing weight sparsity in a delta \tip{rnn}. 
However, it is challenging to implement efficiently because the hardware has to deal with the irregular sparsity pattern in both delta state vectors and weight matrices. 
Previous \tipshort{rnn} hardware accelerators only exploited either spatial sparsity~\cite{han2017ese,Wang2018,Li2019,Cao2019,Wang2019} or temporal sparsity~\cite{GaoDeltaRNN2018,edgedrnn}. 
It is challenging to exploit both sparsity types simultaneously because the hardware has to deal with the static sparsity pattern in weights and the dynamic sparsity pattern in states simultaneously. 
Our paper describes an accelerator that achieves a further speedup of \tip{lstm} \tip{rnn} inference by exploiting spatio-temporal sparsity in both weights and states of the network, as shown in Fig.~\ref{Fig:concept}. The main contributions of this work are:
\begin{enumerate}

\item We extend the delta network algorithm~\cite{neil2016delta} to \tip{lstm} to propose \tip{deltalstm}, inducing temporal sparsity in \tip{lstm} networks (Sec.~\ref{sec:bg}).

      \item 
       We introduce a structured pruning method called \tip{cbtd}\footnote{\url{https://github.com/gaochangw/DeltaLSTM-CBTD}} that produces a balanced workload among columns of a weight matrix (Sec.~\ref{sec:CBTD}). \tip{cbtd} achieves up to 96\% and 94\% weight sparsity of an \tip{lstm} network without accuracy loss, respectively, on the TIMIT and the large-scale Librispeech datasets.
       By comparison with the hardware-optimized \tip{hcgs} pruning method~\cite{Kadetotad2020}, \tip{cbtd} achieves 10$\times$ lower accuracy loss on Librispeech 
       (Sec.~\ref{sec:model_acc}).
       
       \item  We present the first hardware \tip{lstm}-\tipshort{rnn} accelerator, Spartus, that exploits both spatial and temporal sparsity in \tip{lstm} (Sec.~\ref{sec:spartus}). The \tip{lstm} weight matrices are encoded in our customized sparse matrix format called \tip{cbcsc} that can be efficiently processed by the hardware (Sec.~\ref{sec:cbcsc}). 
       \item Spartus is reconfigurable in the number of arithmetic \tips{pe} and thus can be easily implemented on various sizes of \tips{fpga} (Sec.~\ref{sec:mac_array}). Evaluated on the TIMIT dataset, \textbf{Spartus} on the largest Xilinx Zynq \tip{fpga} achieves 1\,$u$s inference latency of an \tip{lstm} layer with 4.7 million parameters, 9.4~TOp/s effective batch-1 throughput, and 1.1~TOp/J effective power efficiency. 
       In comparison to previous \tip{rnn} accelerators (Sect.~\ref{sec:compare_spartus}), the Spartus throughput and power efficiency are respectively 4$\times$ and 7$\times$ higher than previous state-of-the-art \tip{fpga} accelerators BBS~\cite{Cao2019} and DeltaRNN~\cite{GaoDeltaRNN2018} (Sec.~\ref{sec:compare_spartus}). \textbf{Edge-Spartus} on the smallest Zynq \tip{fpga} achieves 33.6\,GOp/s/W effective power efficiency, which is 4$\times$ higher than the previous EdgeDRNN~\cite{edgedrnn} accelerator that only exploited temporal sparsity (Sec.~\ref{sec:compare_edge}), and Edge-Spartus uses inexpensive external memory to run even the largest networks.
\end{enumerate}
\vspace{-2pt}

\section{LSTM \& DeltaLSTM Networks}
\label{sec:bg}

This section introduces the background of the \tip{lstm} networks and describes our proposed variant architecture of \tip{lstm} called \tip{deltalstm} to realize efficient \tip{lstm} inference.
\subsection{LSTM}
An \tip{lstm} unit is composed of an input gate $\mathbf{i}$, a forget gate $\mathbf{f}$, a cell gate $\mathbf{g}$, an output gate $\mathbf{o}$, and a memory cell state $\mathbf{c}$. 
Gates $\mathbf{i}$, $\mathbf{f}$, $\mathbf{g}$ control the update of the cell $\mathbf{c}$ state.
Gate $\mathbf{o}$ determines the proportion of cell memory that is transferred to the hidden state output $\mathbf{h}$. 
In an \tip{lstm} layer, each gate receives two input sequences of length $T$, including an input sequence $X = \{\mathbf{x}_{t}|1\leqslant t\leqslant T, t\in\mathbb{N}\}$ and a sequence of previous hidden states $H_{in} = \{\mathbf{h}_{t}|0\leqslant t\leqslant T-1, t\in\mathbb{N}\}$ from the unit itself. 
At each time step, the \tip{lstm} layer generates a new hidden state vector, giving a sequence $H_{out} = \{\mathbf{h}_{t}|1\leqslant t\leqslant T, t\in\mathbb{N}\}$.

The formulations of an \tip{lstm} layer are given as:
\begin{equation}
\begin{aligned}
	\mathbf{i}_{t} &=\sigma\left(\mathbf{W}_{ii}\mathbf{x}_{t}+\mathbf{b}_{ii}+\mathbf{W}_{hi}\mathbf{h}_{t-1}+\mathbf{b}_{hi}\right) \\
    \mathbf{f}_{t} &=\sigma\left(\mathbf{W}_{if}\mathbf{x}_{t}+\mathbf{b}_{if}+\mathbf{W}_{hf}\mathbf{h}_{t-1}+\mathbf{b}_{hf}\right) \\
    \mathbf{g}_{t} &=\tanh\left(\mathbf{W}_{ig}\mathbf{x}_{t}+\mathbf{b}_{ig}+\mathbf{W}_{hg}\mathbf{h}_{t-1}+\mathbf{b}_{hg}\right) \\
    \mathbf{o}_{t} &=\sigma\left(\mathbf{W}_{io}\mathbf{x}_{t}+\mathbf{b}_{io}+\mathbf{W}_{ho}\mathbf{h}_{t-1}+\mathbf{b}_{ho}\right) \\
    \mathbf{c}_{t} &=\mathbf{f}_{t}\odot \mathbf{c}_{t-1}+\mathbf{i}_{t}\odot \mathbf{g}_{t} \\
    \mathbf{h}_{t} &=\mathbf{o}_{t}\odot \tanh{(\mathbf{c}_{t})}
\end{aligned}
\label{eq:lstm}
\end{equation}
where $W$ denotes weight matrices, $b$ denotes bias vectors, and $\sigma$ denotes the logistic sigmoid function. The symbol $\odot$ signifies pointwise multiplication.

\subsection{DeltaLSTM}
\label{sec:dnalgorithm}
The DeltaLSTM can be understood as follows. Given an input sequence, $X = \{\mathbf{x}_{t}|1\leqslant t\leqslant T, t\in\mathbb{N}\}$, the output is given by $Y = \{\mathbf{y}_{t}|1\leqslant t\leqslant T, t\in\mathbb{N}\}$:
\begin{equation}
\begin{aligned}
  \mathbf{y}_{t}&=\mathbf{W}\mathbf{x}_{t} \\
  \mathbf{y}_{t}&=\mathbf{W}\Delta\mathbf{x}_{t}+\mathbf{y}_{t-1}
  \label{eq:delta}
\end{aligned}
\end{equation}
where \begin{math} \Delta \mathbf{x}_{t} = \mathbf{x}_{t} - \mathbf{x}_{t-1} \end{math} is the difference between the input sequence elements from adjacent time steps and is called a delta vector. $\mathbf{W}$ is the matrix of weight connections from the input to the neurons. The delta vector can be sparse if all its elements below a delta threshold $\Theta$ are set to zero; thus, the term $\mathbf{W}\mathbf{\Delta x}$ in Eq.~(\ref{eq:delta}) becomes a dense matrix - sparse vector multiplication, in which \tip{mac} operations in matrix columns that correspond to zero delta vector elements can be skipped to reduce weight memory access.

The \tip{dn} algorithm was only studied and implemented as \tip{deltagru}.
The \tip{deltalstm} extends the \tip{dn} algorithm to \tip{lstm} \tips{rnn}.
Using \eqref{eq:delta}, the \tip{lstm} equations \eqref{eq:lstm} are converted to the \tip{deltalstm} equations following the formulations in \eqref{eq:deltalstm}:
\begin{equation}
\begin{aligned}
	\mathbf{i}_{t}=\sigma(\mathbf{D}_{i,t}) &= \sigma(\mathbf{W}_{ii}\Delta \mathbf{x}_{t} + \mathbf{W}_{hi}\Delta \mathbf{h}_{t-1} + \mathbf{D}_{i, t-1}) \\
    \mathbf{f}_{t}=\sigma(\mathbf{D}_{f,t}) &= \sigma(\mathbf{W}_{if}\Delta \mathbf{x}_{t} + \mathbf{W}_{hf}\Delta \mathbf{h}_{t-1} + \mathbf{D}_{f, t-1}) \\
    \mathbf{g}_{t}=\tanh(\mathbf{D}_{g,t}) &= \tanh(\mathbf{W}_{ig}\Delta \mathbf{x}_{t} + \mathbf{W}_{hg}\Delta \mathbf{h}_{t-1} + \mathbf{D}_{g, t-1}) \\
    \mathbf{o}_{t} = \sigma(\mathbf{D}_{o,t}) &= \sigma(\mathbf{W}_{io}\Delta \mathbf{x}_{t} + \mathbf{W}_{ho}\Delta \mathbf{h}_{t-1} + \mathbf{D}_{o, t-1}) \\
	\mathbf{c}_{t} &= \mathbf{f}_{t}\odot\mathbf{c}_{t-1} + \mathbf{i}_{t}\odot \mathbf{g}_{t} \\
	\mathbf{h}_{t} &= \mathbf{o}_{t}\odot \tanh\left(\mathbf{c}_{t}\right)
\end{aligned}
\label{eq:deltalstm}
\end{equation}
where the terms $\mathbf{D}$ denote the delta memory for each gate, and they are \tip{mxv} results accumulated over time. 
The delta memory terms in \tip{deltalstm} at $t=1$ correspond to the bias terms in the \tip{lstm} and are initialized to zeros.
Because the delta threshold forces the partial elements of the delta vectors to be zeros, two vectors $\hat{\mathbf{x}}_{t-1}$ and $\hat{\mathbf{h}}_{t-2}$ are used to store the correct previous states to prevent accumulating errors in delta memories. Elements of $\hat{\mathbf{x}}_{t-1}$ and $\hat{\mathbf{h}}_{t-2}$ are updated only when their corresponding delta vector elements are above the delta threshold. The delta vector update process is defined by \eqref{eq:delta_update1}$\sim$\eqref{eq:delta_update4}.
\begin{align}
	\Delta \mathbf{x}_{t}&=\left\{\begin{matrix}
	\mathbf{x}_{t} - \hat{\mathbf{x}}_{t-1}&,\left | \mathbf{x}_{t} - \hat{\mathbf{x}}_{t-1}\right |>\Theta\\ 
	0 & ,\left | \mathbf{x}_{t} - \hat{\mathbf{x}}_{t-1}\right |\leq \Theta
    \end{matrix}\right.  \label{eq:delta_update1} \\
    \hat{\mathbf{x}}_{t-1}&=\left\{\begin{matrix}
    \mathbf{x}_{t-1} & ,\left | \mathbf{x}_{t}-\hat{\mathbf{x}}_{t-1} \right | > \Theta\\ 
    \hat{\mathbf{x}}_{t-2} & , \left | \mathbf{x}_{t}-\hat{\mathbf{x}}_{t-1} \right | \leq \Theta 
    \end{matrix}\right. \label{eq:delta_update2} \\
	\Delta \mathbf{h}_{t-1}&=\left\{\begin{matrix}
	\mathbf{h}_{t-1} - \hat{\mathbf{h}}_{t-2}&,\left | \mathbf{h}_{t-1} - \hat{\mathbf{h}}_{t-2}\right |>\Theta\\ 
	0 & ,\left | \mathbf{h}_{t-1} - \hat{\mathbf{h}}_{t-2}\right |\leq \Theta
    \end{matrix}\right.  \label{eq:delta_update3}\\
    \hat{\mathbf{h}}_{t-2}&=\left\{\begin{matrix}
    \mathbf{h}_{t-2} & ,\left | \mathbf{h}_{t-1}-\hat{\mathbf{h}}_{t-2} \right | > \Theta\\ 
    \hat{\mathbf{h}}_{t-3} & , \left | \mathbf{h}_{t-1}-\hat{\mathbf{h}}_{t-2} \right | \leq \Theta 
    \end{matrix}\right.  \label{eq:delta_update4}
\end{align}
The complete set of \tip{deltalstm} equations is formed by \eqref{eq:deltalstm} and \eqref{eq:delta_update1}$\sim$\eqref{eq:delta_update4}.
\begin{figure}[!t]
	\centering
	\includegraphics[width=0.6\linewidth]{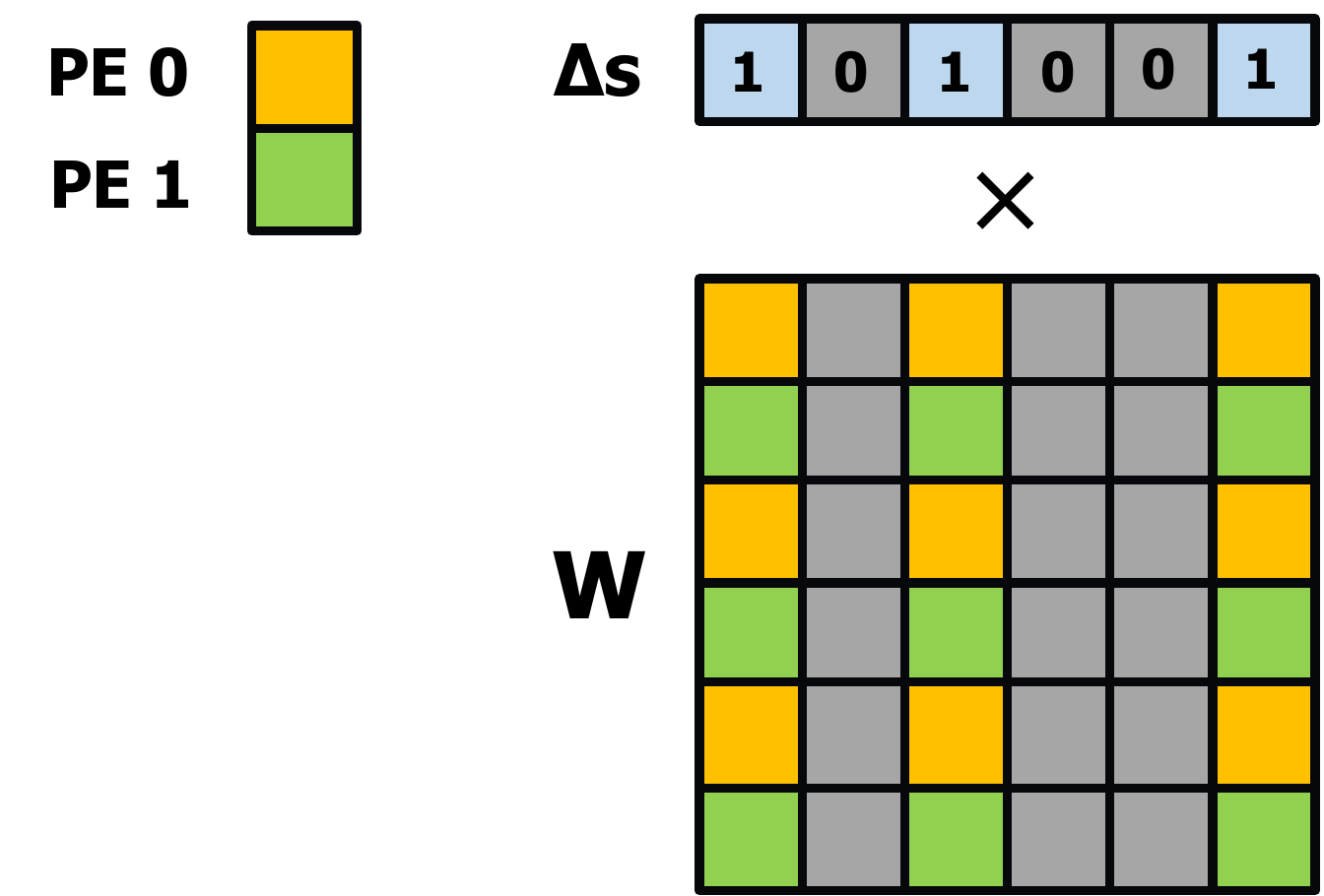}
	\caption{Example of matrix-vector multiplication in Spartus between the weight matrix, $\mathbf{W}$, and the delta state vector, $\Delta s$. The workload assignment of 2 \tips{pe} is shown here. Skipped columns due to zeros in $\Delta s$ are highlighted in gray.}
	\label{Fig:pe_mxv}
\end{figure}
\section{Structured Pruning of DeltaLSTM}
\label{sec:StPrune}
The \tip{dn} algorithm converts dense-matrix-dense-vector multiplication in \tip{lstm} to dense-matrix-sparse-vector multiplication to save columns of \tip{mxv} operations, but the remaining columns are still dense.
These dense columns can be sparsified by pruning. 
However, random access of weight matrix columns is required to exploit temporal sparsity in delta state vectors and this leads to difficulties in further exploiting sparsity in weight columns; 
that is, parallelizing \tip{mac} operations for hardware \tips{pe} on remaining nonzero elements. 

Although fine-grain pruning methods~\cite{Han2015} could achieve high sparsity around 90\% in \tipshort{rnn} weights with negligible accuracy loss, they introduce irregular nonzero element distribution in the pruned sparse matrix  leading to an unbalanced workload for \tips{pe}. Other structured pruning methods like BBS can create structured sparse weights, but the workload will not be balanced once combined with temporal \tip{deltalstm}. This is because the nonzero values are not evenly distributed across columns and will lead to an unbalanced workload when combined with temporal sparsity. 
In this work, we propose a hardware-oriented pruning method called \tip{cbtd} that balances the workload even with  the random weight column access in \tip{deltalstm}.

\subsection{Column-Balanced Targeted Dropout (CBTD)}
\label{sec:CBTD}

\begin{algorithm}[!htb]
\SetAlgoLined
\KwData{
    $\mathbf{A}$, matrix; \\
    \hspace{1cm} $Q$, the number of columns in $\mathbf{A}$; \\
    \hspace{1cm} $H$, the height of columns in $\mathbf{A}$; \\
    \hspace{1cm} $\gamma$, target sparsity; \\
    \hspace{1cm} $\alpha$, dropout probability; \\
    \hspace{1cm} $M$, number of \tips{pe} allocated along a column.
}
\KwResult{A sparse weight matrix $\mathbf{B}$ of which columns having balanced workload for each \tip{pe}\;}
Build set $C$ containing all columns of $\mathbf{A}$, where $C=\{\mathbf{c}_j| \mathbf{c}_j\in \mathbb{R}^{H}, 1\leqslant j \leqslant Q, j\in\mathbb{N}\}$; \\
 Shuffle and split columns in set $C$ into subcolumns $S=\{\mathbf{s}_{ij}|\mathbf{s}_{ij}\in\mathbb{R}^{H/M}, 1\leqslant i \leqslant M, 1\leqslant j \leqslant Q, i,j\in \mathbb{N}\}$ \;
\For{$j=1$ to $Q$}{
 
 \For{$i=1$ to $M$}{
  Sort elements of $\mathbf{s}_{ij}$ by their magnitudes\;
  Set the smallest $\floor*{H/M*\gamma}$ elements in $\mathbf{s}_{ij}$\ to zero with a probability of $\alpha$;
 }
}
Reverse the shuffling and splitting to build a sparse matrix $\mathbf{B}$ from subcolumns in set $S$\;
\Return{B}
 \caption{\small{\glsreset{cbtd}\tip{cbtd}}}
\end{algorithm}
\begin{algorithm}[!t]
\SetAlgoLined
\KwData{
    $\mathbf{W}$, \tip{lstm} weight matrices; \\
    \hspace{1cm} $Q$, the number of columns in $\mathbf{W}$; \\
    \hspace{1cm} $H$, the height of columns in $\mathbf{W}$; \\
    \hspace{1cm} $\gamma$, target sparsity; \\
    \hspace{1cm} $\alpha$, dropout probability; \\
    \hspace{1cm} $\Delta \alpha$, step size of dropout probability; \\
    \hspace{1cm} $M$, number of \tips{pe} per column.
}
\KwResult{Trained network with a sparse weight matrix in which columns have a balanced workload for each \tip{pe}\;}
 $\alpha=0$\;
  \For{iterations}{
   Forward Propagation\;
   Backward Propagation\;
   Update Parameters $\mathbf{W}$\;
   CBTD ($\mathbf{W}$, $Q$, $H$, $\gamma$, $\alpha$, $M$)\;
   \If{$\alpha<1$}{
    $\alpha=\alpha+\Delta \alpha$\;
   }
 }
 \caption{\tip{lstm} Training with \tip{cbtd}}
\end{algorithm}
\begin{figure*}[!t]
	\centering
	\includegraphics[width=0.6\linewidth]{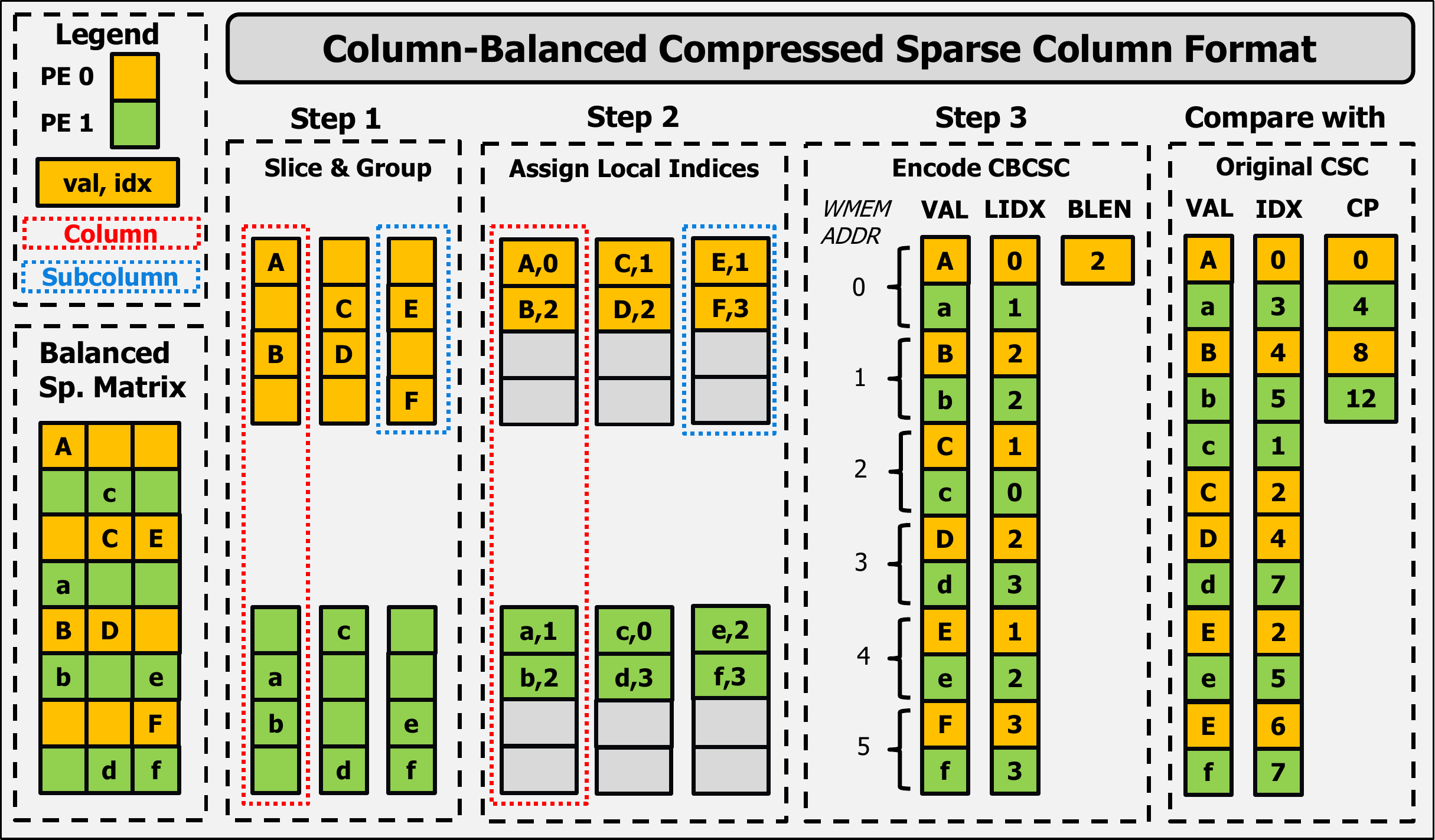}
	\caption{\glsreset{cbcsc}Steps to encode a sparse matrix into the \tip{cbcsc} format}
	\label{Fig:cbcsc}
\end{figure*}

\begin{algorithm}[!t]
\SetAlgoLined
\KwData{
                 $\mathbf{A}$, a matrix; \\
    \hspace{1cm} $Q$, the number of columns in $\mathbf{A}$; \\
    \hspace{1cm} $H$, the height of columns in $\mathbf{A}$; \\
    \hspace{1cm} $M$, the number of \tips{pe} in a \tip{mac} array; \\
    \hspace{1cm} $\gamma$, target sparsity; \\
        }
\KwResult{Sparse matrix in \tip{cbcsc} format stored in \tip{val}, \textbf{\tip{lidx}}, \tip{blen}\;}
Build set $C$ containing all columns of $\mathbf{A}$, where $C=\{\mathbf{c}_j| \mathbf{c}_j\in \mathbb{R}^{H}, 1\leqslant j \leqslant Q, j\in\mathbb{N}\}$; \\
 Shuffle and split columns in set $C$ into subcolumns $S\{\mathbf{s}_{ij}|\mathbf{s}_j\in\mathbb{R}^{H/M}, 1\leqslant i \leqslant M, 1\leqslant j \leqslant Q, i,j\in \mathbb{N}\}$\;
 \For{$j=1$ to $Q$}{
  \par
  \For{$i=1$ to $M$}{
   \par
   \For{$k=1$ to $H/M$}{ \par
    \If{$s_{ij}[k]\neq0$}{
        $\mathbf{\tip{val}}$.append($s_{ij}[k]$); \par 
        $\mathbf{\tip{lidx}}$.append($k$);
    }
   }
  }
 }
 $\mathbf{\tip{blen}} = \ceil*{H/M*(1-\gamma)}$; \par
 \caption{\tip{cbcsc} Format Encoding}
\end{algorithm}
As shown in Fig.~\ref{Fig:pe_mxv}, interleaved rows of \tip{mxv} workload are assigned to \tips{pe} that have \tip{mac} units for \tip{mxv} computation in the Spartus accelerator. 
The procedure of applying \tip{cbtd} on a weight matrix is shown in Algorithm 1. 
Given $M$, which is the number of \tips{pe} along the column direction in the Spartus accelerator, \tip{cbtd} splits each column into the same number of groups (also called subcolumns) as the number of \tips{pe}.
Thus, $M$ determines the size of each subcolumn for an \tip{lstm} layer of a certain size.
Next, weight elements in each subcolumn are sorted by their magnitudes. 
Then, the smallest $\floor*{H/M*\gamma}$ portion of elements in each subcolumn is set to zero with a dropout probability $\alpha$ and a target sparsity, 
$\gamma$. The $\alpha$ probability was used to introduce stochasticity in the targeted dropout process.
The same $\gamma$ and $\alpha$ are used for all subcolumns to ensure the same number of nonzero elements in each subcolumn.
Therefore, $M$ determines the granularity of \tip{cbtd}. 
With a larger $M$, the locations of nonzero weights are more tightly constrained due to the smaller subcolumn size and vice versa.

\subsection{LSTM Training with CBTD}

\tip{cbtd} is applied to the training procedure of \tip{lstm} networks as described in Algorithm 2. 
The \tip{cbtd} is used to set relatively unimportant weights to zeros in each epoch after the parameter update step. 
Weights that are set to zero in the previous epoch are allowed to recover during the parameter update step in the next epoch.
The target sparsity $\gamma$ is fixed throughout the training process. 
The dropout probability, $\alpha$, is increased gradually from 0 to 1 with a step size of $\Delta \alpha$, which determines the number of epochs needed for the \tip{lstm} weight sparsity to reach the target sparsity $\gamma$.
This training method guarantees that the network reaches the target sparsity and the same number of nonzero elements between columns or between subcolumns at the end of the training.

\subsection{Column-Balanced Compressed Sparse Column Format}
\label{sec:cbcsc}

To fully utilize weight sparsity in an \tipshort{rnn} pruned by \tip{cbtd}, we propose a new sparse weight matrix format method called \tip{cbcsc} based on the original \tip{csc} format~\cite{Duff1989}. 
A sparse matrix encoded in \tip{csc} has 3 vectors: 
\tip{val} for nonzero weight elements, \tip{idx} containing the indices of elements in \tip{val}, and \tip{cp} containing pointers to the start of a new column. 
The problem of \tip{csc} is that the nonzero elements are not arranged in a regular form that benefits \tip{pe} access. During run-time, the number of weight elements for each \tip{pe} at the memory interface is different, requiring arbitration between \tips{pe} to ensure the correct dispatching of weights. 
Arbitration reduces the effective memory bandwidth for weight access.
To overcome this problem, we propose \tip{cbcsc} to force the same number of weight elements for each \tip{pe} at the memory interface.
The procedure of \tip{cbcsc} encoding is illustrated in Fig.~\ref{Fig:cbcsc} and the steps are described below. 
\begin{enumerate}
    \item Assign interleaved rows to \tips{pe}. Columns of the weight matrix are sliced, and interleaved elements in each column are grouped into subcolumns. Each subcolumn is assigned to a single \tip{pe}.
    \item Find the local index of each nonzero element within the subcolumn it resides in. Zero elements in each subcolumn are discarded, and nonzero elements are aggregated into a dense vector.
    \item Encode \tip{cbcsc} by allocating nonzero values and corresponding indices into two vectors \tip{val}, \tip{lidx}. Another scalar value called \tip{blen} indicates the number of nonzero elements in each subcolumn. This process is described in Algorithm 3.
\end{enumerate}

In contrast to \tip{csc}, the weight matrix encoded by \tip{cbcsc} has the same number of nonzero elements in \tip{val} for each \tip{pe}; thus, arbitration is not needed when fetching weight data and the logic area can be reduced.
\section{Accelerator Design}
\label{sec:spartus}
\subsection{Top-level Architecture}
\begin{figure*}[!t]
	\centering
	\includegraphics[width=0.5\linewidth]{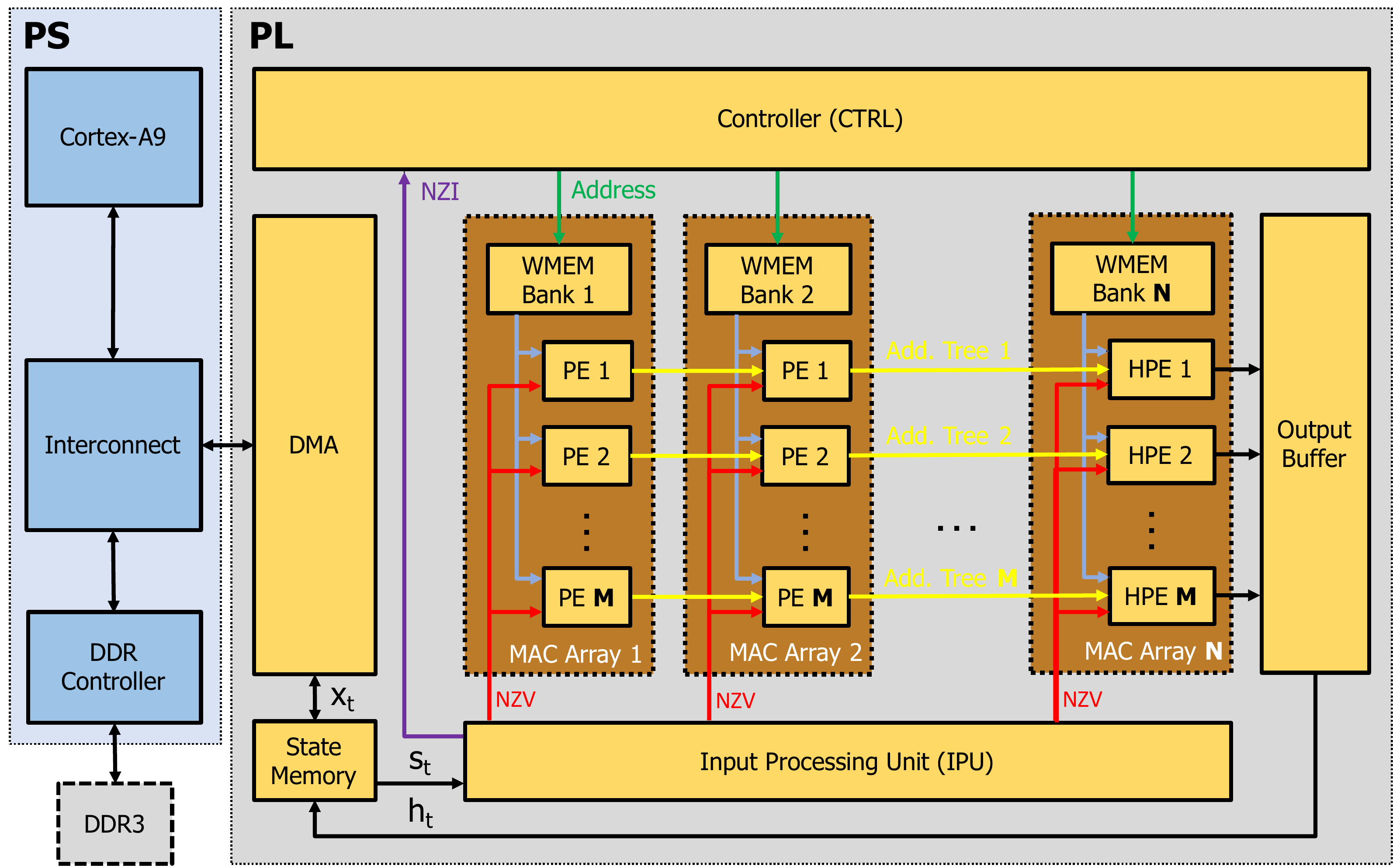}
	\caption{Spartus accelerator with reconfigurable $\boldsymbol{N}$ \tip{mac} arrays,  and  $\boldsymbol{M}$ \tips{pe} in each array.}
	\label{Fig:system}
\end{figure*}

Fig.~\ref{Fig:system} shows the top-level architecture of the Spartus accelerator. The accelerator was implemented on the \tip{pl} of Xilinx Zynq \tip{soc}, which also has an ARM Cortex-A9 CPU as the host on the \tip{ps} side.
Spartus is composed of a \tip{ctrl}, an \tip{smem} block, an \tip{ipu}, \tip{mac} arrays, \tip{wmem} banks, \tips{at}, and an \tip{ob}. An Xilinx \tip{dma} IP block controlled by the host is used to manage I/O communications between the accelerator and the host.
Input vectors, $\mathbf{x}_{t}$, are streamed from \tip{ps} to \tip{pl} through the \tip{dma} module and buffered in the \tip{smem} block to hide the transfer latency. The \tip{smem} block is also used to buffer \tip{lstm} activations $\mathbf{h}_{t}$. 
The \tip{ipu} concatenates $\mathbf{x}_{t}$ and $\mathbf{h}_{t}$ to compute the delta state vectors $\Delta s_{t}$. 
The delta state vectors are encoded into \tips{nzv} and \tips{nzi}. 
\tips{nzi} are dispatched to \tip{ctrl} to generate the physical memory addresses of weights in each \tip{wmem} bank, and \tips{nzv} are dispatched to \tip{mac} arrays to be multiplied with the fetched weights.
There are $N$ \tip{mac} arrays. 
Each of the first $N-1$ arrays contains $M$ \tips{pe} that perform \tip{mxv} between \tips{nzv} and corresponding weight columns. 
The last array has $M$ \tips{hpe} that are also responsible for post-\tip{mxv} activation generation. 
The \tip{ob} helps to hide the latency of transferring the last layer's activations to the host.

\begin{figure}[!t]
	\centering
	\includegraphics[width=0.6\linewidth]{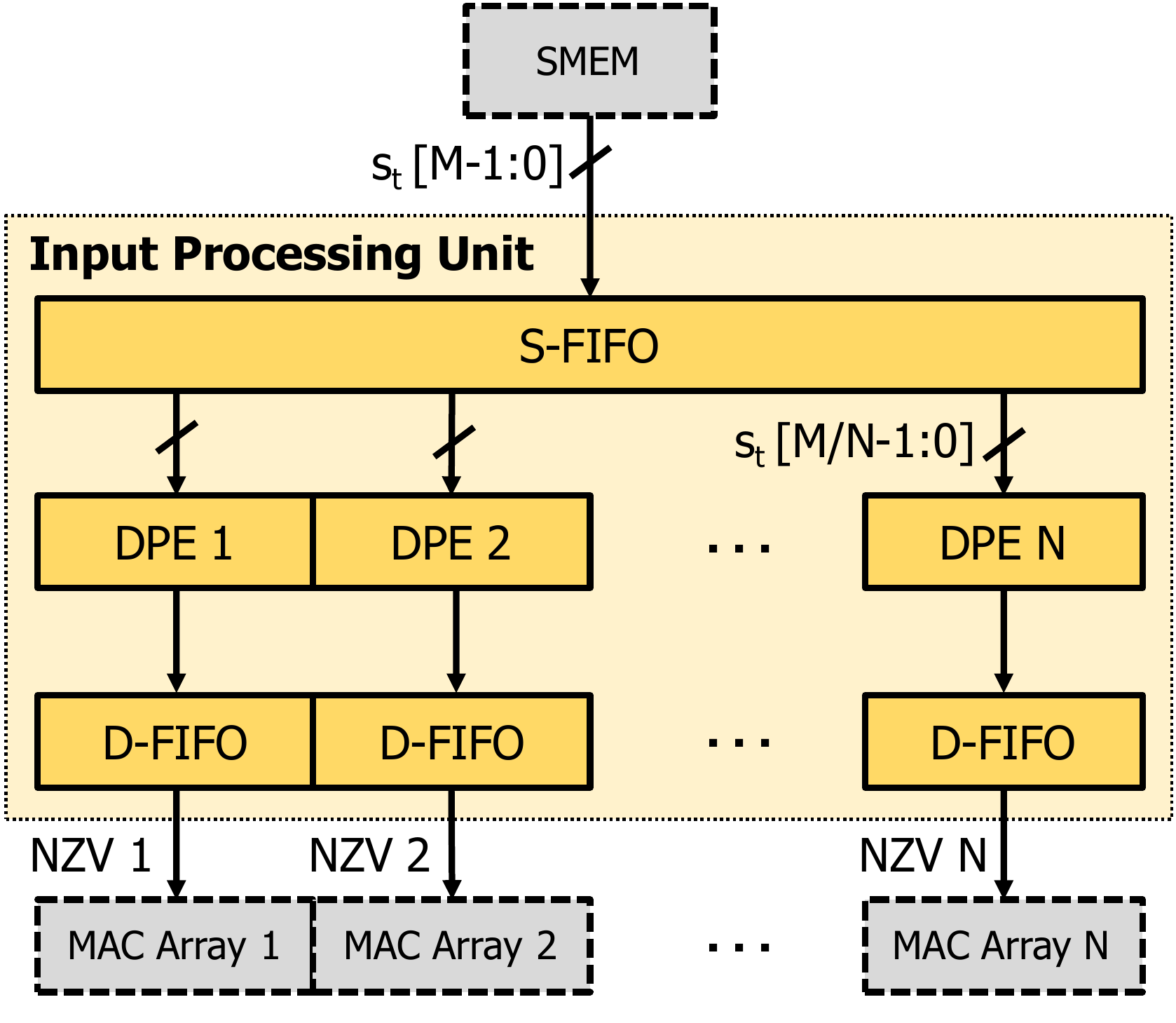}
	\caption{Structure of the \tip{ipu}}
	\label{Fig:ipu}
\end{figure}
\begin{figure}[!t]
	\centering
	\includegraphics[width=0.7\linewidth]{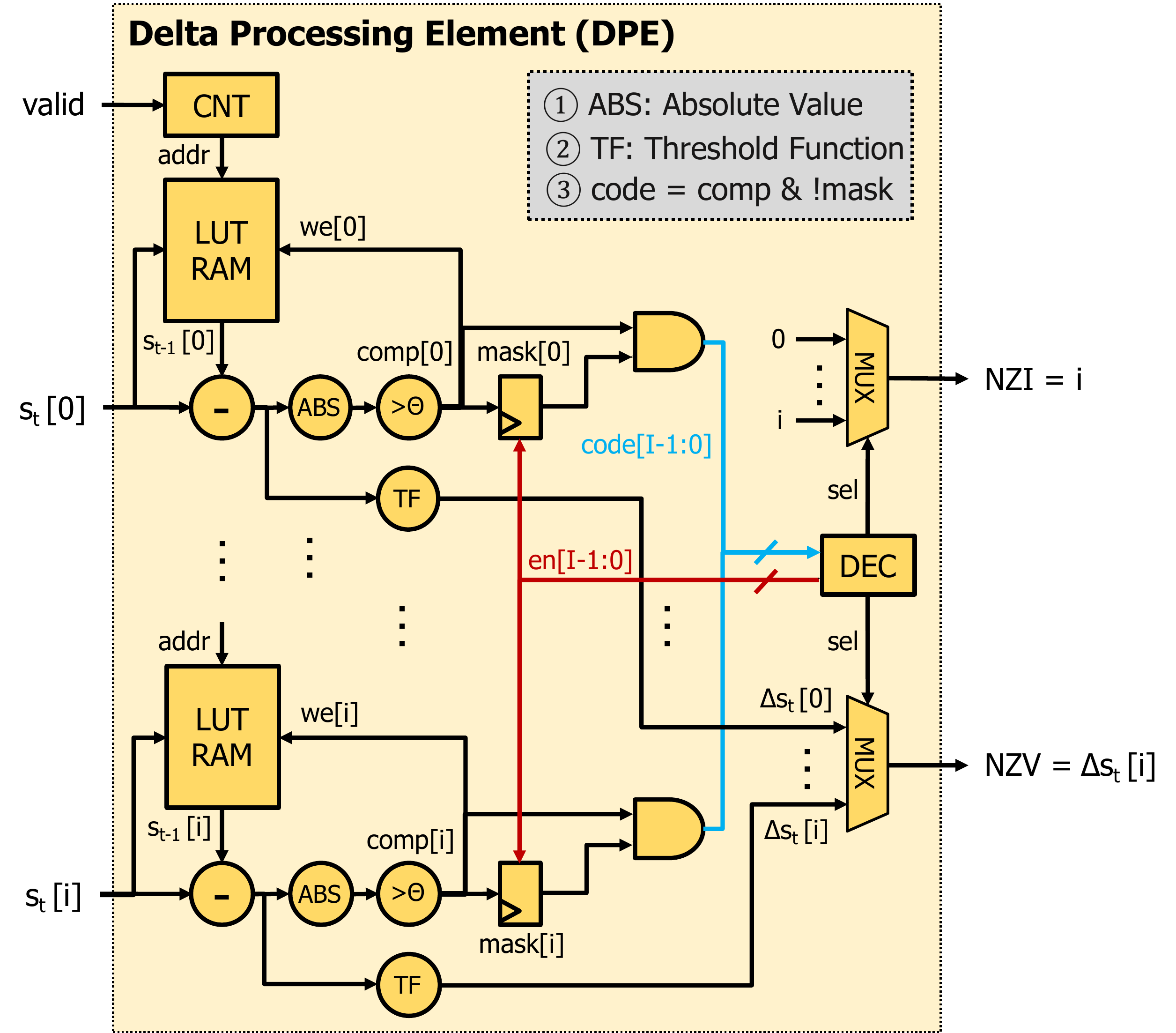}
	\caption{Simplified data flow of a \tip{dpe}}
	\label{Fig:dpe}
\end{figure}
\begin{figure}[!t]
	\centering
	\includegraphics[width=0.7\linewidth]{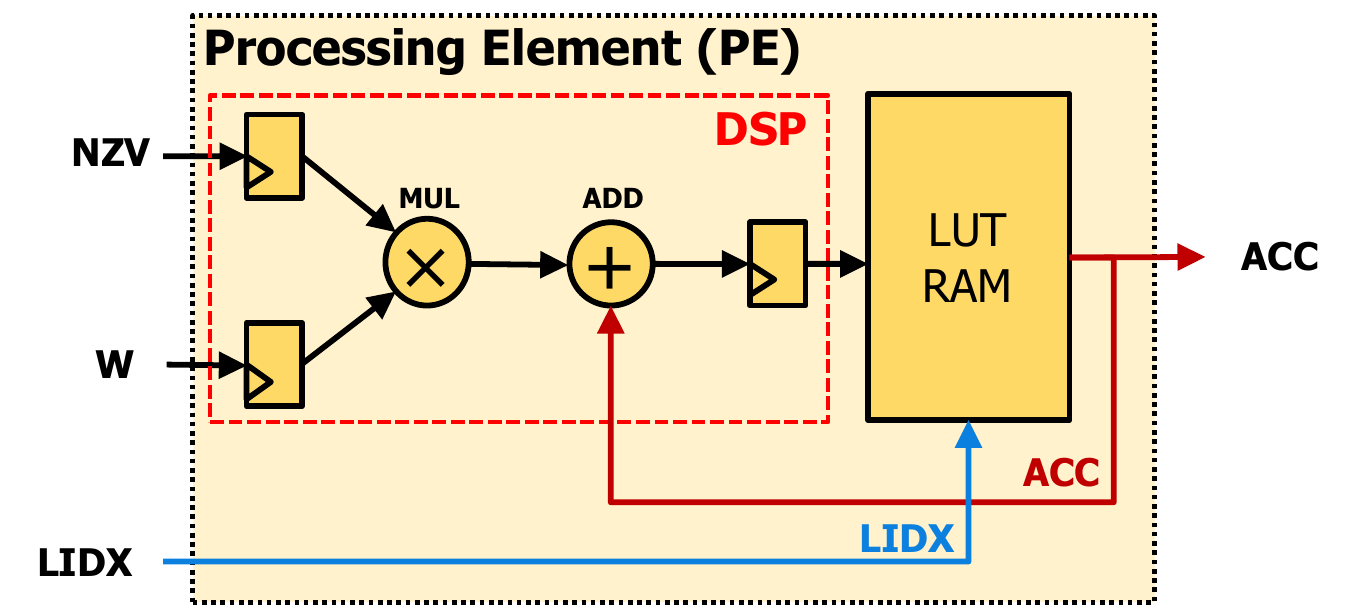}
	\caption{Architecture of a \tip{pe}}
	\label{Fig:pe}
\end{figure}
\begin{figure}[!t]
	\centering
	\includegraphics[width=0.7\linewidth]{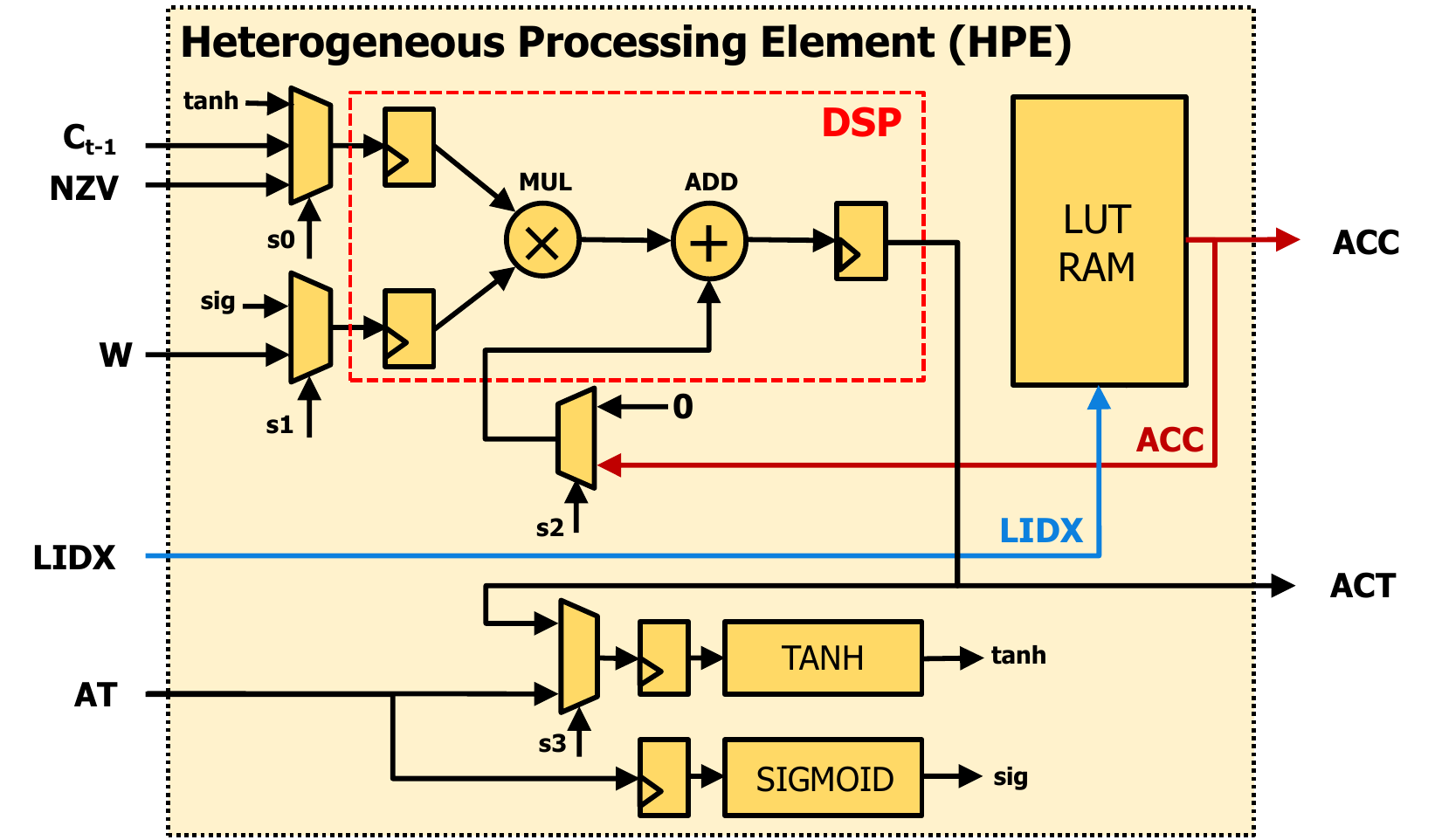}
	\caption{Architecture of a \tip{hpe}}
	\label{Fig:hpe}
\end{figure}
\begin{figure*}[!t]
	\centering
	\includegraphics[width=0.7\linewidth]{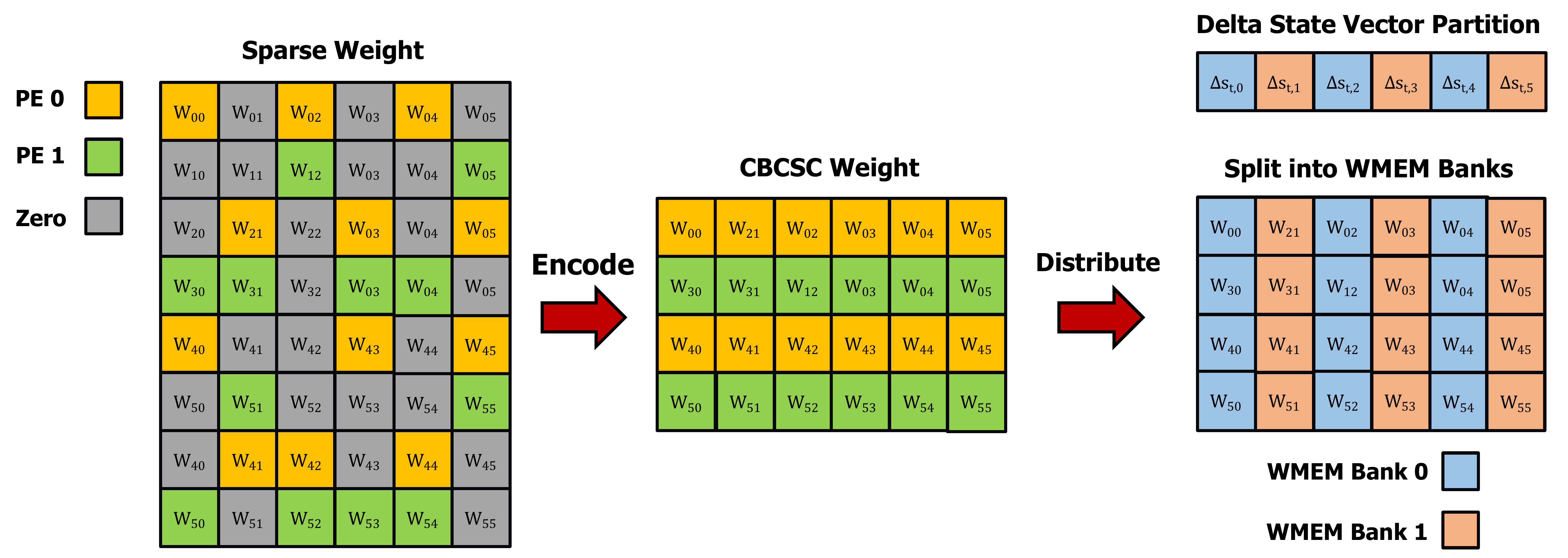}
	\caption{Sparse \tip{deltalstm} weights are encoded into \tip{cbcsc} format and then split into \tip{wmem} banks for each \tip{mac} array to access independently. We show an example with $N=2$ \tip{mac} arrays and $M=2$ \tips{pe}/\tips{hpe} per array.}
	\label{Fig:weight}
\end{figure*}
\subsection{Input Processing Unit (\textbf{\tip{ipu}})}
The \tip{ipu} computes delta state vectors $\Delta s_{t}$ from both input vectors $\mathbf{x}_{t}$ and hidden layer activations $\mathbf{h}_{t}$. It then generates \tip{nzv} and \tip{nzi} that respectively contain the nonzero values of $\Delta s_{t}$ and their corresponding indices. 
As shown in Fig~\ref{Fig:ipu}, inputs of the \tip{ipu} are streamed in from the \tip{smem} block. 
In \tip{smem}, the lengths of $\mathbf{x}_{t}$ and $\mathbf{h}_{t}$ vectors are zero-padded to the length that is a multiple of $M$ and concatenated into a single state vector $\mathbf{s}_{t}$. 
During the inference, the \tip{ipu} receives $M$ elements of $\mathbf{s}_{t}$ per clock cycle whenever the state \tip{fifo} (S-\tip{fifo}) is not full. 
The state vector $\mathbf{s}_{t}[M-1:0]$ is then partitioned into $N$ equal segments $\mathbf{s}_{t}[M/N-1:0]$, each of which is fed into a \tip{dpe} shown in Fig.~\ref{Fig:dpe}.
Details of the state vector partition will be discussed in Section V-E. 
\subsection{Delta Processing Unit (\textbf{\tip{dpe}})}
Fig.~\ref{Fig:dpe} shows the architecture of the \tip{dpe}.
Following \eqref{eq:delta_update1}$\sim$\eqref{eq:delta_update4}, the \tip{dpe} calculates a delta vector $\Delta s_{t}$ from $s_{t}$ and $s_{t-1}$.
Each \tip{dpe} receives a partitioned $s_{t}$ segment of length $I = M/N$, given that $M$ must be divisible by $N$.
The input elements of a \tip{dpe} are denoted as $S = \{s_{t}[i]|0\leqslant i\leqslant I-1, i\in\mathbb{N}\}$.
$s_{t}$ is buffered in the \tip{ipu} input \tip{fifo} and $s_{t-1}$ is stored in the look-up table-based memory (\textbf{\tip{lutram}}) blocks in the \tip{dpe}.
Each \tip{lutram} block is addressed by a \tip{cnt}, which is incremented by one when $valid$ is asserted.
\eqref{eq:delta_update1} and \eqref{eq:delta_update3} are implemented as the \tip{tf} block while \eqref{eq:delta_update2} and \eqref{eq:delta_update4} are realized by controlling the write enable ($we[i]$) signal of the \tip{lutram} using the output of a comparator that produces a high logic state when $|\Delta \mathbf{s}_{t}|$ is larger than the delta threshold $\Theta$.
Next, the first nonzero element in $\Delta \mathbf{s}_{t}$ with its index $i$ are selected as \tip{nzv} and \tip{nzi} respectively, by two multiplexers controlled by a \tip{dec} according to signal \textit{code}[M/N-1:0]. 
$\mathit{code}=\mathit{comp}\,\&\,\mathit{mask}$, where elements of $\mathit{mask}$ are initialized as ones. 
Controlled by the signal $en[i]$, when $\mathit{mask}[i]==1$ and $comp[i]==1$ in the current cycle, $\mathit{mask}[i]$ is overridden by zero in the next clock cycle.
In this way, $\mathit{mask}$ is used to disable nonzero $\Delta s_{t}$ once it is already selected and added to the \tip{nzv}. 
The \tips{nzv} and \tips{nzi} generated from \tips{dpe} are buffered in their corresponding delta state \tip{fifo} (D-\tip{fifo}), which drives the input of \tips{pe} or \tips{hpe} in \tip{mac} arrays.
\subsection{Multiply-Accumulate (MAC) Arrays}
\label{sec:mac_array}
\tip{spmxspv} in Spartus are handled by the $N$ \tip{mac} arrays. Each array receives \tips{nzv} from its corresponding \tip{dpe} in the \tip{ipu}. 
In Fig.~\ref{Fig:system}, the leftmost $N-1$ \tip{mac} arrays have $M$ \tips{pe} and the rightmost \tip{mac} array has $M$ \tips{hpe}. 
As shown in Fig.~\ref{Fig:pe}, a \tip{pe} has a \tip{mac} unit synthesized by a \tip{dsp} block that performs up to 16-bit by 16-bit multiplication and 48-bit accumulation between \tips{nzv} and weights. 
In addition to the \tip{dsp} block, the \tip{hpe} shown in Fig.~\ref{Fig:hpe} has multiplexers before each input operand of the \tip{dsp} to reuse for point-wise multiplication and addition. Furthermore, the \tip{hpe} also has \textsl{tanh} and \textsl{sigmoid} blocks implemented by look-up tables.

Each \tip{pe} or \tip{hpe} has a dedicated \tip{lutram} block to buffer their corresponding partial sums. The \tip{lutram} is addressed by the IDX of weights encoded in the \tip{cbcsc} format.
\begin{table*}[!t]
\centering
\caption{Comparison of \tip{fpga} chips ordered according to their on-chip memory size in Megabits (Mb).}
\label{tab:fpga}
\begin{threeparttable}
\begin{tabular}{cccccccc}

\textbf{FPGA}      & \textbf{DSP} & \textbf{\tnote{1}~BRAM/M20K} & \textbf{\tnote{2}~LUT/ALM} & \textbf{FF} & \textbf{Process} & \textbf{\tnote{3}~Cost} & \textbf{\tipshort{rnn} Accelerator}        \\  \hline

\textbf{XC7Z007S}  & 66           & 50 (1.8 Mb)        & 14,400           & 28,800      & 28nm             & \$55          & EdgeDRNN~\cite{edgedrnn}, \textbf{Edge-Spartus} \\ 
\textbf{XC7Z100}   & 2,020        & 755 (26.5 Mb)      & 277,400          & 554,800     & 28nm             & \$3,192       & DeltaRNN~\cite{GaoDeltaRNN2018}, \textbf{Spartus}   \\  \hline

{XCKU060}   & 2,760        & 1,080 (38.9 Mb)    & 331,680          & 663,360     & 20nm             & \$3,978       & \tip{ese}~\cite{han2017ese}                     \\ 
{SX660}     & 3,376        & 2,133 (41.7 Mb)    & 250,540          & 1,002,160   & 20nm             & N/A           & E-\tip{lstm}~\cite{Wang2019}                  \\ 
{XC7VX690T} & 3,600        & 1,470 (51.7 Mb)    & 433,200          & 866,400     & 28nm             & \$17,926      & C-\tip{lstm}~\cite{Wang2018}, E-\tipshort{rnn}~\cite{Li2019}           \\ 
{GX1150}    & 3,036        & 2,713 (53.0 Mb)    & 427,200          & 1,708,800   & 20nm             & N/A           & \tip{bbs}~\cite{Cao2019}                     \\ \hline
\end{tabular}
\begin{tablenotes}\footnotesize
\item[1] Single BRAM Size = 36 Kb. Single M20K size = 20 Kb. 
\item[2] The \tip{lut} in Xilinx FPGAs~\cite{xilinxwhite} is not equivalent to the \tip{alm} in Intel FPGAs~\cite{intelwhite}.
\item[3] Costs extracted from Digi-Key US in March 2022.

\end{tablenotes}
\end{threeparttable}
\end{table*}

After SpMxSpV, delta memory terms $DM$, defined in \eqref{eq:deltalstm}, are obtained by accumulating partial sums of all \tip{mac} arrays by $M$ adder trees. Outputs of the adder trees are then fed to \tips{hpe} throughput the port \textit{AT} for activation generation. Activation $h_{t}$ are first stored in \tip{smem} for delta state vector computation of the next time step and then streamed out to the host through \tip{dma}.

\subsection{Network Adaptation}
\label{sec:na}
The Spartus accelerator supports fixed-point weights and activations. 
To run a \tip{deltalstm} network on Spartus, the network should be trained starting with floating-point parameters and quantized to fixed-point numbers before inference. To run an \tip{lstm} network with Spartus, weight matrices $W_{ii}$, $W_{hi}$, $W_{if}$, $W_{hf}$, $W_{ig}$, $W_{hg}$, $W_{io}$, $W_{ho}$ are stacked to a single matrix $W_{s}$ given by \eqref{eq:weight}.
\begin{equation}
W_{s} =
\begin{pmatrix}
W_{ii} & W_{hi}\\ 
W_{ig} & W_{hg}\\ 
W_{if} & W_{hf}\\ 
W_{io} & W_{ho}
\end{pmatrix}
    \label{eq:weight}
\end{equation}

During matrix-vector multiplication, the stacked weight matrix is multiplied by the delta state vector $\mathbf{\Delta s}_{t}$, which follows the same partition pattern as the state vector $\mathbf{s}_{t}$.
Accordingly, $W_{s}$ is partitioned into $N$ submatrices. 
Each submatrix only contains columns that will be multiplied by \tips{nzv} from its corresponding \tip{dpe} and are encoded into the \tip{cbcsc} format, as shown in Fig.~\ref{Fig:weight} and discussed in Section~\ref{sec:cbcsc}. 
Then, the interleaved columns of the \tip{cbcsc} weight are split into $N$ submatrices. 
Each submatrix is stored in a \tip{wmem} bank dedicated to its corresponding \tip{mac} array to avoid bank conflict. 
The delta vector partition pattern follows how the stacked weight matrix is split.
The network weights are quantized to 8 bits, while 8 and 10-bit \tip{lidx} are used for Spartus and Edge-Spartus, respectively, in the \tip{cbcsc} format. 

\section{Experimental Setup}
\subsection{Hardware Implementation}
\begin{figure}[t]
	\centering
	\includegraphics[width=0.55\linewidth]{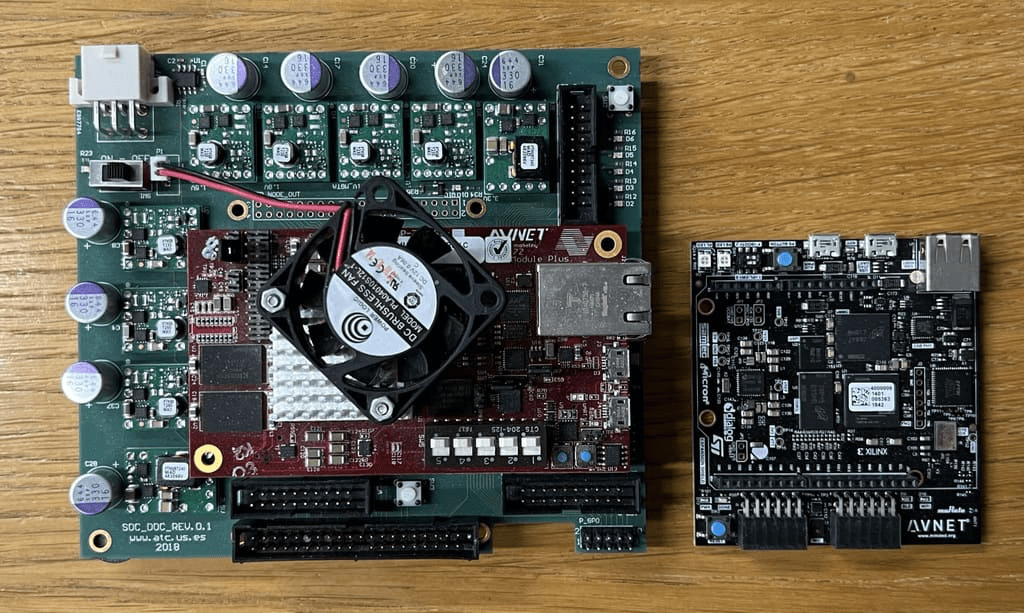}
	\caption{\tip{fpga} development boards used in this work. (Left) AVNET Zynq XC7Z100 with a custom baseboard for Spartus; (Right) AVNET MiniZed XC7Z007S for Edge-Spartus. See Table~\ref{tab:fpga}.}
	\label{fig:board}
\end{figure}
To demonstrate the scalability of Spartus, we evaluated its hardware performance on two configurations with different numbers of \tips{pe}, \textbf{Spartus} and \textbf{Edge-Spartus}, both of which are synthesized in Vivado 2018.2. 
Spartus is implemented on an AVNET Zynq \tip{mmp} \tip{som} that has a Xilinx Zynq XC7Z100 \tip{soc}, the largest Zynq \tip{soc} on the market. 
Edge-Spartus is implemented on an AVNET MiniZed development board having a Zynq XC7007S \tip{soc}, the smallest Zynq \tip{soc} on the market. 
Photos of the boards are shown in Fig.~\ref{fig:board}.
Both \tips{soc} have an ARM Cortex-A9 \tip{cpu} in their \tip{ps} as the host.
Table~\ref{tab:fpga} compares the amount of available logic and memory resources and the cost of both Zynq \tips{soc} with those used by other \tipshort{rnn} accelerators.

The Spartus accelerator resides on the \tip{pl} side of both \tips{soc} and is controlled by bare-metal C programs compiled in the Xilinx \tip{sdk}. 
The Spartus accelerator buffers network weights using on-chip \tip{bram} and the C program initializes \tip{lstm} weights in the DDR3 \tip{dram} and then streams weights to the on-chip \tip{wmem} banks synthesized with \tips{bram}. 
In Edge-Spartus, the on-chip \tip{bram} capacity is too small to buffer the large network; thus the program initializes weights in off-chip DDR3L \tip{dram} and the accelerator fetches weights from the \tip{hp} AXI4 slave port interfacing \tip{pl} and \tip{ps}.

\subsection{Feature Extraction \& Network Setup}
We evaluate the impact on the accuracy of both the \tip{cbtd} method and \tip{deltalstm} algorithm on a speech recognition task using the TIMIT and larger Librispeech datasets.

The TIMIT dataset~\cite{timit} has a standard 462 speaker training set with all dialect sentences (SA) removed following~\cite{Graves2013}. 
The \tip{dev} and \tip{test} sets have utterances of 50 and 24 speakers respectively.
123-dimensional features are extracted from the input and consist of the 40 coefficients of FFT-based filter banks distributed on a Mel-scale, plus the energy term; and their first and second-order temporal derivatives~\cite{Graves2013}. 
The \tip{ctc} loss function~\cite{graves2006connectionist} is used during training so that the final logit layer of the \tip{am} generates phonemes directly without requiring a sophisticated decoding algorithm. The \tip{per} results were collected using a simple greedy decoder that selects the index of the phoneme class with the highest score.

The large-scale LibriSpeech dataset has 1000 hours of audiobook speech based on LibriVox's library~\cite{Panayotov2015}. For Librispeech, we followed the same \tip{am} training process leading to the \tip{lstm} baseline in the PyTorch-Kaldi framework using the 100-hour subset~\cite{Ravanelli2019}. 
We used the Kaldi toolkit~\cite{povey2011kaldi} to extract 40-dimensional \tip{fmllr} features~\cite{GALES199875} and decode the \tip{am} output using a Viterbi decoder with a 4-gram language model to calculate the \tip{wer}. The PyTorch-Kaldi~\cite{Ravanelli2019} framework was used to train the \tip{am}. 
We used this training and evaluation process to have a fair comparison with the PyTorch-Kaldi baseline and the \tip{hcgs}~\cite{Kadetotad2020} method.

We trained \tip{lstm}-\tips{am} on both datasets. 
The last \tip{lstm} layer is followed by a \tip{fcl} having the same number of units and a final logit layer. The networks implemented on the Spartus hardware accelerators are quantized during training to 8-bit weights and 16-bit activations using the dual-copy rounding method~\cite{Stromatias2015}.
\begin{figure*}[!t]
	\centering
	\includegraphics[width=0.85\linewidth]{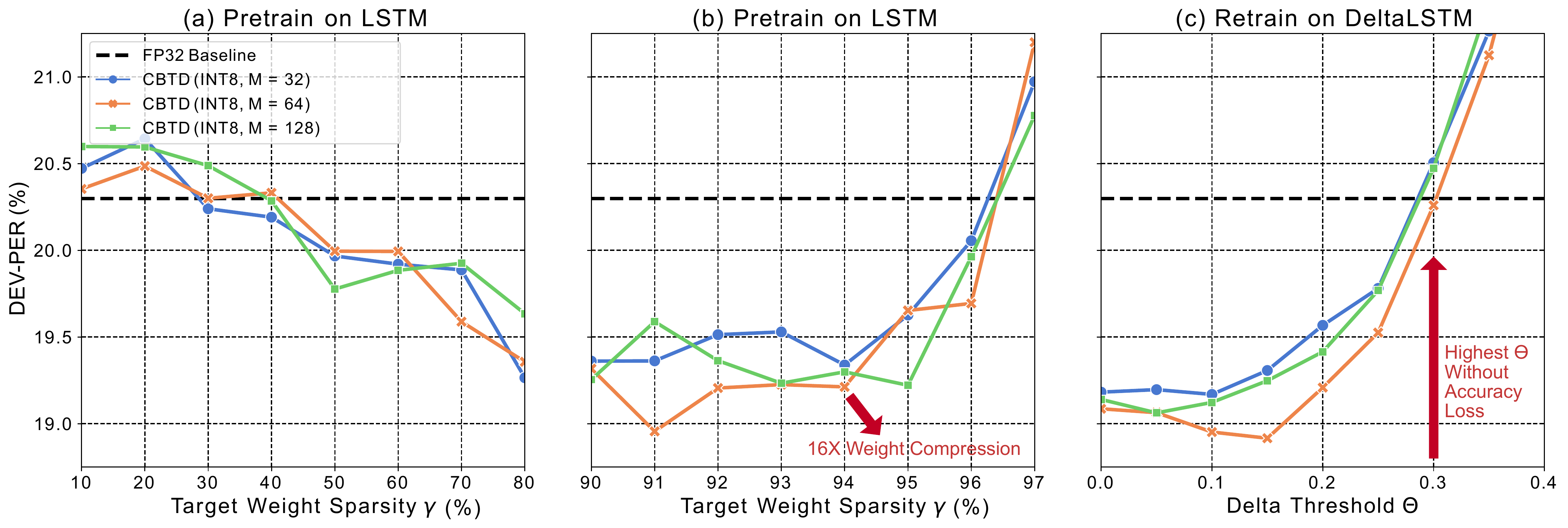}
	\caption{
	\tip{per} evaluated on the TIMIT development set (\tip{dev}-\tip{per}). (a) and (b) show the dependence of \tip{per} on different target weight sparsity values after the 150-epoch pretrain phase on \tip{lstm}.
	\textbf{(a)} \tip{per} from $\gamma=10\%$ to $\gamma=80\%$ (step size =\,10\%). \textbf{(b)} \tip{per} from $\gamma=90\%$ to $\gamma=97\%$ (step size =\,1\%). \textbf{(c)}   \tip{per} after the 50-epoch retrain phase on \tip{deltalstm} for different delta thresholds. The pretrained models with $\gamma=94\%$ from (b) were used in the retrain phase. Results were averaged from 5 runs.
	}
	\label{Fig:Explore_Array}
\end{figure*}
\begin{figure}[!t]
	\centering
	\includegraphics[width=0.6\linewidth]{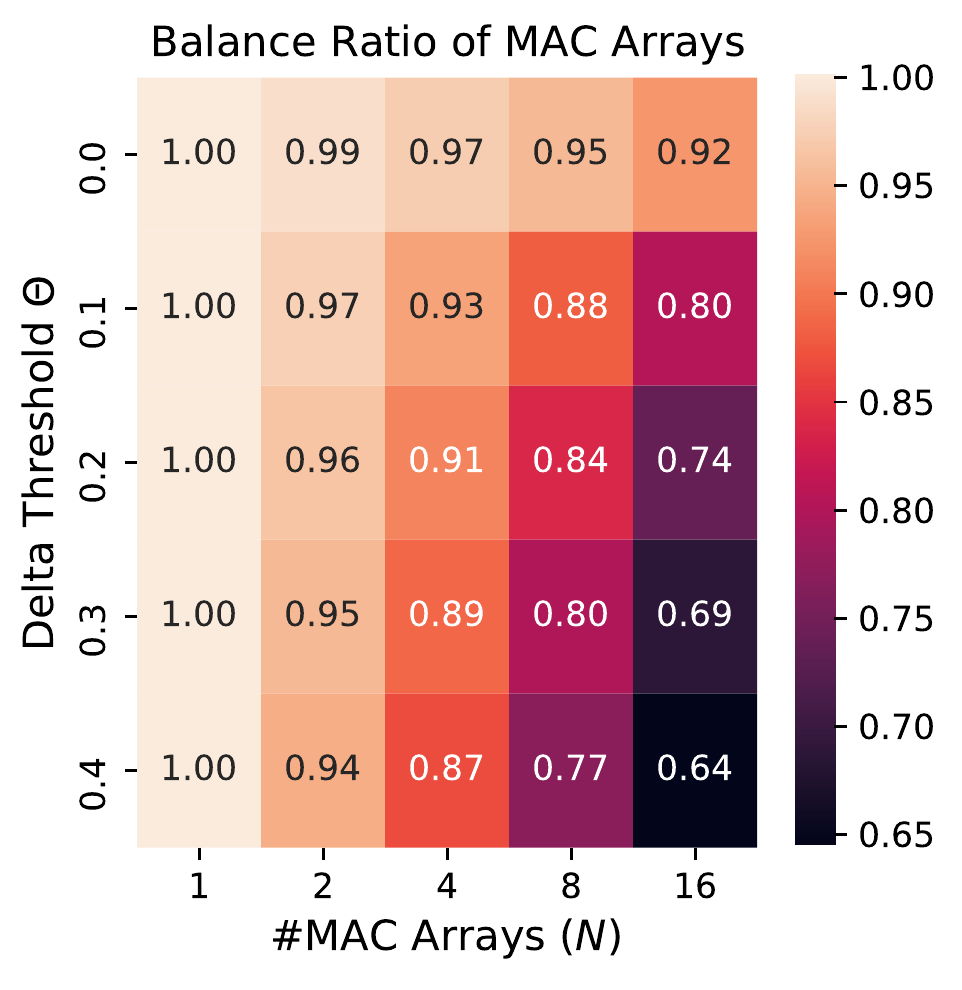}
	\caption{Balance ratio of \tip{mac} arrays with respect to delta threshold and  the number of \tip{mac} arrays in the Spartus accelerator. The values were obtained by running the hardware over all TIMIT core test set samples.}
	\label{Fig:balance}
\end{figure}

\subsection{Pretrain \& Retrain}
We adopted a pretrain-retrain process for this task to train networks with spatio-temporal sparsity.
For the TIMIT dataset, during the pretrain phase, the \tip{lstm} layers were initialized and trained with the \tip{cbtd} applied for 150 epochs. 
We set the step size of dropout probability $\Delta \alpha$ to 1/30 so that the target sparsity $\gamma$ was achieved within the first 30 epochs and maintained during the remaining epochs.
For the Librispeech dataset, we pretrain the networks for 24 epochs (the same as the PyTorch Kaldi baseline) and anneal the step size of dropout probability of \tip{cbtd} during the first 10 epochs.
The \tip{cbtd} was also applied to the \tip{fcl} with the same $\Delta \alpha$ and $\gamma$.
The pretraining process was early stopped at the epoch that achieved the best accuracy on the development set. 

During the retraining phase, weights of the pretrained \tip{lstm} layers are copied into \tip{deltalstm} layers of the same size to be retrained for 50 epochs for the TIMIT dataset and 5 epochs for the Librispeech dataset with $\alpha=1$ . 
The retrain phase was also early stopped at the best accuracy on the development set. 
The final accuracy results for the TIMIT dataset are reported on the core test set, which has 192 samples. For Librispeech, the final \tip{wer} results were evaluated on the standard test set.

\section{Results}
\label{sec:results}
\subsection{Design Space Exploration}
We determine how the number of \tips{pe}, and \tip{mac} arrays; and the delta threshold $\Theta$, impact the accuracy and hardware performance as evaluated on TIMIT.

\subsubsection{Pretrain: Phone Error Rate vs. Weight Sparsity}
This section reports the impact of \tip{cbtd} on weight sparsity and accuracy.
Fig.~\ref{Fig:Explore_Array}(a) \& Fig.~\ref{Fig:Explore_Array}(b) show the evolution of \tip{per} on the development set (\tip{dev}-\tip{per}) over the target sparsity. 
Each data point in Fig.~\ref{Fig:Explore_Array} is the average value over 5 runs. 
We explored using different numbers  $M$ of \tips{pe} per \tip{mac} array, with values of $M$=32/64/128 for the \tip{cbtd} method.
The horizontal dashed line shows the \tip{dev}-\tip{per} of the network with 32-bit floating-point (FP32) parameters trained without using \tip{cbtd} and quantization, which achieved 20.30\% \tip{dev}-\tip{per}.

Results show that the accuracy with different numbers of \tips{pe} is similar. 
Most of the \tip{dev}-\tip{per} values of \tip{lstm} networks trained with \tip{cbtd} were worse than the FP32 baseline when the target sparsity is below 40\%. 
Similar to the Dropout~\cite{Srivastava2014} and Targeted Dropout~\cite{Gomez2018} regularization methods, \tip{cbtd} helped regularize the \tip{lstm} network. 
Target sparsity $\gamma$ between 50\% and 90\% resulted in better \tip{dev}-\tip{per} than FP32 due to the regularization effect of \tip{cbtd}. 
A similar regularization effect on the TIMIT dataset due to weight pruning on \tip{lstm} was also observed in~\cite{Cao2019}.
The best \tip{per} results are achieved when $\gamma$ is between 90\% and 94\%.
The pretrained networks with \tip{cbtd} achieve up to 96\% weight sparsity without loss of accuracy when compared to the FP32 baseline results. 
To achieve the best \tip{per} while having speedup as high as possible, the network having 94\% weights sparsity was used in the retrain phase, which achieves a \tip{dev}-\tip{per} value of around 19.30\% among the best \tip{per} values of all sparsity levels.

\subsubsection{Retrain: Phone Error Rate vs. Delta Threshold}
The purpose of the retrain phase is to induce temporal sparsity using the \tip{deltalstm} model.
Temporal sparsity introduces zero elements that can be skipped to save the computations and the memory access but can lead to accuracy degradation once the sparsity is higher than a certain value.
This section reports the results of the retrain phase and explores the optimal condition to achieve a balance between speedup and accuracy.

Fig.~\ref{Fig:Explore_Array}(c) shows the \tip{dev}-\tip{per} of the network retrained with \tip{deltalstm} layers. 
During the retraining phase, various delta thresholds $\Theta$ were used. The same thresholds were applied to all \tip{lstm} layers. 
The \tip{cbtd} was still used during the whole training phase with a step size of dropout rate $\alpha$ fixed to 1.
Results show that \tip{dev}-\tip{per} increases with increased delta thresholds $\Theta$ and the network trained with $M=64$ achieves the best \tip{dev}-\tip{per} which is lower  than the FP32 baseline until $\Theta > 0.3$. 
Thus, we set $\Theta=0.3$ to evaluate the performance of the accelerator in Sec.~\ref{sec:results}.

\begin{table*}[!t]
\centering
\caption{Accuracy results on the core test set of TIMIT with different sizes of \tip{lstm} networks and optimization methods. (\textbf{L} denotes the number of \tip{lstm} layers; \textbf{H} denotes the number of units in each \tip{lstm} layer; \textbf{UNI} denotes that the network is unidirectional; the bold network names are supported by Spartus, and the bold rows have the highest arithmetic operations saving without losing accuracy; results were averaged from 5 runs.)}
\label{tab:timit}
\resizebox{0.98\textwidth}{!}{ 
\begin{tabular}{|l|cc|cccc|cc|c|}
\hline
\multicolumn{1}{|c|}{\textbf{Network}}                                                                                        & \multicolumn{1}{c|}{\textbf{$\gamma$}} & \textbf{$\Theta$} & \multicolumn{1}{c|}{\textbf{\begin{tabular}[c]{@{}c@{}}Weight \\ Precision\end{tabular}}} & \multicolumn{1}{c|}{\textbf{\begin{tabular}[c]{@{}c@{}}Model Size \\ (MB)\end{tabular}}} & \multicolumn{1}{c|}{\textbf{\begin{tabular}[c]{@{}c@{}}Weight \\ Sparsity\\ (\%)\end{tabular}}} & \textbf{\begin{tabular}[c]{@{}c@{}}Temporal \\ Sparsity\\ (\%)\end{tabular}} & \multicolumn{1}{c|}{\textbf{\begin{tabular}[c]{@{}c@{}}TEST-PER\\ (\%)\end{tabular}}} & \textbf{\begin{tabular}[c]{@{}c@{}}Accuracy\\ Improvement\\ (\%)\end{tabular}} & \textbf{\begin{tabular}[c]{@{}c@{}}Arithmetic\\ Operations\\ Saving\end{tabular}} \\ \hline
\multirow{2}{*}{\tip{lstm}-3L-512H-UNI}                                                                                             & -                                      & -                 & FP32                                                                                      & 23.18                                                                                    & 0                                                                                               & 0                                                                            & 22.7$\pm$0.16                                                                         & 0                                                                              & 1$\times$                                                                         \\
                                                                                                                              & -                                      & -                 & INT8                                                                                      & 5.79                                                                                     & 0                                                                                               & 0                                                                            & 22.8$\pm$0.50                                                                         & -0.1                                                                           & 1$\times$                                                                         \\ \cline{1-1}
\multirow{4}{*}{\textbf{\begin{tabular}[c]{@{}l@{}}\tip{lstm}-3L-512H-UNI-CBTD\\ (w/ Spatial Sparsity)\end{tabular}}}               & 0.80                                   & -                 & INT8                                                                                      & 1.27                                                                                     & 78.13                                                                                           & 0                                                                            & 21.5$\pm$0.12                                                                         & 1.3                                                                            & 4.6$\times$                                                                       \\
                                                                                                                              & 0.90                                   & -                 & INT8                                                                                      & 0.72                                                                                     & 87.50                                                                                           & 0                                                                            & 21.3$\pm$0.54                                                                         & 1.4                                                                            & 8.0$\times$                                                                       \\
                                                                                                                              & \textbf{0.94}                          & \textbf{-}        & \textbf{INT8}                                                                             & \textbf{0.36}                                                                            & \textbf{93.75}                                                                                  & \textbf{0}                                                                   & \textbf{22.6$\pm$0.47}                                                                & \textbf{0.2}                                                                   & \textbf{16.0$\times$}                                                             \\
                                                                                                                              & 0.97                                   & -                 & INT8                                                                                      & 0.18                                                                                     & 96.88                                                                                           & 0                                                                            & 25.7$\pm$0.56                                                                         & -2.9                                                                           & 32.1$\times$                                                                      \\ \hline
\multirow{2}{*}{\tip{lstm}-2L-768H-UNI}                                                                                             & -                                      & -                 & FP32                                                                                      & 30.88                                                                                    & 0                                                                                               & 0                                                                            & 23.2$\pm$0.36                                                                         & 0                                                                              & 1$\times$                                                                         \\
                                                                                                                              & -                                      & -                 & INT8                                                                                      & 8.10                                                                                     & 0                                                                                               & 0                                                                            & 23.4$\pm$0.31                                                                         & 0.2                                                                            & 1$\times$                                                                         \\ \cline{1-1}
\multirow{4}{*}{\textbf{\begin{tabular}[c]{@{}l@{}}\tip{lstm}-2L-768H-UNI-CBTD\\ (w/ Spatial Sparsity)\end{tabular}}}               & 0.80                                   & -                 & INT8                                                                                      & 1.69                                                                                     & 79.17                                                                                           & 0                                                                            & 21.7$\pm$0.27                                                                         & 1.5                                                                            & 4.8$\times$                                                                       \\
                                                                                                                              & 0.90                                   & -                 & INT8                                                                                      & 0.84                                                                                     & 89.58                                                                                           & 0                                                                            & 21.4$\pm$0.15                                                                         & 2.0                                                                            & 9.6$\times$                                                                       \\
                                                                                                                              & 0.94                                   & -                 & INT8                                                                                      & 0.51                                                                                     & 93.75                                                                                           & 0                                                                            & 22.4$\pm$0.96                                                                         & 0.8                                                                            & 16.0$\times$                                                                      \\
                                                                                                                              & \textbf{0.97}                          & \textbf{-}        & \textbf{INT8}                                                                             & \textbf{0.34}                                                                            & \textbf{95.83}                                                                                  & \textbf{0}                                                                   & \textbf{22.5$\pm$0.32}                                                                & \textbf{0.7}                                                                   & \textbf{24.0$\times$}                                                             \\ \hline
\multirow{2}{*}{\tip{lstm}-2L-1024H-UNI}                                                                                            & -                                      & -                 & FP32                                                                                      & 56.81                                                                                    & 0                                                                                               & 0                                                                            & 22.3$\pm$0.29                                                                         & 0                                                                              & 1$\times$                                                                         \\
                                                                                                                              & -                                      & -                 & INT8                                                                                      & 14.20                                                                                    & 0                                                                                               & 0                                                                            & 22.0$\pm$0.29                                                                         & 0.3                                                                            & 1$\times$                                                                         \\ \cline{1-1}
\multirow{4}{*}{\textbf{\begin{tabular}[c]{@{}l@{}}\tip{lstm}-2L-1024H-UNI-CBTD\\ (w/ Spatial Sparsity)\end{tabular}}}              & 0.80                                   & -                 & INT8                                                                                      & 2.88                                                                                     & 79.69                                                                                           & 0                                                                            & 21.1$\pm$0.31                                                                         & 0.9                                                                            & 4.9$\times$                                                                       \\
                                                                                                                              & 0.90                                   & -                 & INT8                                                                                      & 1.55                                                                                     & 89.06                                                                                           & 0                                                                            & 20.8$\pm$0.33                                                                         & 1.5                                                                            & 9.1$\times$                                                                       \\
                                                                                                                              & \textbf{0.94}                          & \textbf{-}        & \textbf{INT8}                                                                             & \textbf{0.89}                                                                            & \textbf{93.75}                                                                                  & \textbf{0}                                                                   & \textbf{20.6$\pm$0.27}                                                                & \textbf{1.7}                                                                   & \textbf{16.0$\times$}                                                             \\
                                                                                                                              & 0.97                                   & -                 & INT8                                                                                      & 0.44                                                                                     & 96.88                                                                                           & 0                                                                            & 22.7$\pm$0.39                                                                         & -0.4                                                                           & 32.1$\times$                                                                      \\ \cline{1-1}
\multirow{2}{*}{\textbf{\begin{tabular}[c]{@{}l@{}}\tip{deltalstm}-2L-1024H-UNI-CBTD\\ (w/ Spatio-Temporal Sparsity)\end{tabular}}} & 0.94                                   & 0.1               & INT8                                                                                      & 0.89                                                                                     & 93.75                                                                                           & 74.22                                                                        & 20.6$\pm$0.30                                                                         & 1.7                                                                            & 62.1$\times$                                                                      \\
                                                                                                                              & \textbf{0.94}                          & \textbf{0.3}      & \textbf{INT8}                                                                             & \textbf{0.89}                                                                            & \textbf{93.75}                                                                                  & \textbf{90.60}                                                               & \textbf{21.8$\pm$0.30}                                                                & \textbf{0.6}                                                                   & \textbf{170.2$\times$}                                                            \\ \hline
\end{tabular}
}
\end{table*}
\begin{table*}[!t]
\centering
\caption{Accuracy results on the test set of Librispeech with different sizes of \tip{lstm} networks and optimization methods trained on the 100-hour training set. (\textbf{BI} denotes that the network is bidirectional; Accuracy Improvement is the \tip{wer} difference to the baseline floating-point conventional \tipshort{rnn}.)}
\label{tab:librispeech}
\resizebox{0.98\textwidth}{!}{ 
\begin{tabular}{|l|cc|cccc|cc|c|}
\hline
\multicolumn{1}{|c|}{\textbf{Network}}                                                                                        & \multicolumn{1}{c|}{\textbf{$\gamma$}} & \textbf{$\Theta$} & \multicolumn{1}{c|}{\textbf{\begin{tabular}[c]{@{}c@{}}Weight \\ Precision\end{tabular}}} & \multicolumn{1}{c|}{\textbf{\begin{tabular}[c]{@{}c@{}}Model Size \\ (MB)\end{tabular}}} & \multicolumn{1}{c|}{\textbf{\begin{tabular}[c]{@{}c@{}}Weight \\ Sparsity \\ (\%)\end{tabular}}} & \textbf{\begin{tabular}[c]{@{}c@{}}Temporal \\ Sparsity \\ (\%)\end{tabular}} & \multicolumn{1}{c|}{\textbf{\begin{tabular}[c]{@{}c@{}}TEST-WER\\ (\%)\end{tabular}}} & \textbf{\begin{tabular}[c]{@{}c@{}}Accuracy\\ Improvement\\ (\%)\end{tabular}} & \textbf{\begin{tabular}[c]{@{}c@{}}Arithmetic\\ Operations\\ Saving\end{tabular}} \\ \hline
\begin{tabular}[c]{@{}l@{}}\tip{lstm}-4L-512H-BI \\ (PyTorch-Kaldi Baseline~\cite{Ravanelli2019})\end{tabular}                                           & -                                      & -                 & FP32                                                                                      & 14.84                                                                                    & 0                                                                                                & 0                                                                             & 6.4                                                                                   & -                                                                              & 1$\times$                                                                         \\ \cline{1-1}
\multirow{2}{*}{\textbf{\begin{tabular}[c]{@{}l@{}}\tip{lstm}-4L-512H-BI-CBTD\\ (w/ Spatial Sparsity)\end{tabular}}}                & \textbf{0.90}                          & \textbf{-}        & \textbf{INT8}                                                                             & \textbf{0.46}                                                                            & \textbf{87.5}                                                                                    & \textbf{0}                                                                    & \textbf{6.2}                                                                          & \textbf{0.2}                                                                   & \textbf{8.0$\times$}                                                              \\
                                                                                                                              & 0.94                                   & -                 & INT8                                                                                      & 0.23                                                                                     & 93.75                                                                                            & 0                                                                             & 7.1                                                                                   & -0.7                                                                           & 16.0$\times$                                                                      \\ \hline
\tip{lstm}-3L-512H-UNI                                                                                                              & -                                      & -                 & FP32                                                                                      & 21.3                                                                                     & 0                                                                                                & 0                                                                             & 7.0                                                                                   & -                                                                              & 1$\times$                                                                         \\ \cline{1-1}
\begin{tabular}[c]{@{}l@{}}\tip{lstm}-3L-512H-UNI-HCGS~\cite{Kadetotad2020}\\ (w/ Spatial Sparsity)\end{tabular}                                         & -                                      & -                 & INT6                                                                                      & 0.29                                                                                     & 93.75                                                                                            & 0                                                                             & 11.4                                                                                  & -4.4                                                                              & 16.0$\times$                                                                      \\ \cline{1-1}
\textbf{\begin{tabular}[c]{@{}l@{}}\tip{lstm}-3L-512H-UNI-CBTD\\ (w/ Spatial Sparsity)\end{tabular}}                                & 0.94                                   & -                 & INT8                                                                                      & 0.33                                                                                     & 93.75                                                                                            & 0                                                                             & 7.3                                                                                   & -0.3                                                                           & 16.0$\times$                                                                      \\ \hline
\multirow{2}{*}{\tip{lstm}-2L-1024H-UNI}                                                                                            & -                                      & -                 & FP32                                                                                      & 56.81                                                                                    & 0                                                                                                & 0                                                                             & 8.0                                                                                   & -                                                                              & 1$\times$                                                                         \\
                                                                                                                              & -                                      & -                 & INT8                                                                                      & 14.20                                                                                    & 0                                                                                                & 0                                                                             & 7.6                                                                                   & 0.4                                                                            & 1$\times$                                                                         \\ \cline{1-1}
\multirow{4}{*}{\textbf{\begin{tabular}[c]{@{}l@{}}\tip{lstm}-2L-1024H-UNI-CBTD\\ (w/ Spatial Sparsity)\end{tabular}}}              & 0.80                                   & -                 & INT8                                                                                      & 2.88                                                                                     & 79.69                                                                                            & 0                                                                             & 6.7                                                                                   & 1.3                                                                            & 4.9$\times$                                                                       \\
                                                                                                                              & 0.90                                   & -                 & INT8                                                                                      & 1.55                                                                                     & 89.06                                                                                            & 0                                                                             & 6.5                                                                                   & 1.5                                                                            & 9.1$\times$                                                                       \\
                                                                                                                              & \textbf{0.94}                          & \textbf{-}        & \textbf{INT8}                                                                             & \textbf{0.89}                                                                            & \textbf{93.75}                                                                                   & \textbf{0}                                                                    & \textbf{7.0}                                                                          & \textbf{1.0}                                                                   & \textbf{16.0$\times$}                                                             \\
                                                                                                                              & 0.97                                   & -                 & INT8                                                                                      & 0.44                                                                                     & 96.88                                                                                            & 0                                                                             & 8.3                                                                                   & -0.3                                                                           & 32.1$\times$                                                                      \\ \cline{1-1}
\multirow{2}{*}{\textbf{\begin{tabular}[c]{@{}l@{}}\tip{deltalstm}-2L-1024H-UNI-CBTD\\ (w/ Spatio-Temporal Sparsity)\end{tabular}}} & \textbf{0.94}                          & \textbf{0.05}     & \textbf{INT8}                                                                             & \textbf{0.89}                                                                            & \textbf{93.75}                                                                                   & \textbf{80.45}                                                                & \textbf{7.4}                                                                          & \textbf{0.6}                                                                   & \textbf{81.8$\times$}                                                             \\
                                                                                                                              & 0.94                                   & 0.1               & INT8                                                                                      & 0.89                                                                                     & 93.75                                                                                            & 89.91                                                                         & 8.5                                                                                   & -0.5                                                                           & 158.6$\times$                                                                     \\ \hline
\end{tabular}
}
\end{table*}

\subsubsection{Workload Balance Between MAC Arrays}
The theoretical peak throughput $\nu_{Peak}$ of the accelerator on the \tip{pl} is given as:
\begin{equation}
\nu_{Peak} = 2\cdot f_{pl}\cdot K
\label{eq:peak_throughput}
\end{equation}
where $f_{pl}$ is the operation clock frequency of the \tip{pl} and $K$ is the total number of \tip{mac} units. In Spartus, $K = M\times N$, where $N$ is the number of \tip{mac} arrays. 
Thus, the theoretical peak throughput is proportional to the number of \tip{mac} arrays; however, the actual hardware throughput is affected by the workload imbalance between \tip{mac} arrays.

Unlike the structured sparse weight matrix induced by \tip{cbtd}, the sparsity pattern of the delta state vector $s_{t}$ is dynamically updated during \tip{lstm} inference in each time step. 
Therefore, partitioned $\Delta s_{t}$ vector segments, which are encoded as \tips{nzv} by \tips{dpe}, are likely to have different numbers of nonzero values in each time step of the \tip{lstm} computation. 
Provided that there are $N$ \tip{mac} arrays in the accelerator, the workload (WL), which is the number of nonzero elements of $\Delta s_{t}$ allocated to the $n$-th \tip{mac} array at time step $t$, is $\mathbf{WL}_{t}^{n}$. Then, the \tip{br} of \tip{mac} arrays is given as:
\begin{equation}
\begin{aligned}
    \mathbf{BR} &= \frac{\sum_{t=1}^{T}\mathbf{WL}_{t, mean}}{\sum_{t=1}^{T}\mathbf{WL}_{t, max}} \\
    \mathbf{WL}_{t, mean} &= \frac{\sum_{n=1}^{N}\mathbf{WL}_{t}^{n}}{N} \\
    \mathbf{WL}_{t, max} &= max\left(\mathbf{WL}_{t}^{1},\mathbf{WL}_{t}^{2},...,\mathbf{WL}_{t}^{N} \right) \\
\end{aligned}
\label{eq:workload}
\end{equation}

\noindent where $\mathbf{WL}_{t, mean}$ and $\mathbf{WL}_{t, max}$ are respectively the mean and max workload of all \tip{mac} arrays at time step $t$. 
The \tip{br} is obtained by running the hardware on a temporal sequence with a length of $T$ and getting the sum of $\mathbf{WL}_{t, mean}$ and $\mathbf{WL}_{t max}$ over $T$ time steps. 
The performance of hardware is bottlenecked by the max workload $\mathbf{WL}_{t, max}$ at each time $t$.
The optimal workload balance can be achieved once the workload of each \tip{mac} array equals to $\mathbf{WL}_{t, mean}$.
Therefore, the closer \tip{br}  is to 1, the more balanced the workload between \tip{mac} arrays.

Fig.~\ref{Fig:balance} shows the \tip{br} values of \tip{mac} arrays evaluated on all samples in the core test set of TIMIT running the best \tip{deltalstm} network selected after the retraining phase. 
Increasing delta threshold $\Theta$ or the number of \tips{pe} resulted in a more imbalanced allocation of nonzero elements between \tip{mac} arrays. 
To maximize the hardware performance at $\Theta=0.3$, we set the number of \tip{mac} arrays between $N=8$ and $N=16$, where \tip{br} is between 0.8 and 0.69. 
Thus, the loss of hardware performance due to workload imbalance is 20\% to 31\%, which is small compared to the 8$\times$ to 16$\times$ peak throughput gain by increasing $N$ from 1 to 8 and 16. 
However, using $N>8$ resulted in routing congestion. 
\tip{fpga}
Therefore, we set $M=64$ \tip{mac} arrays and $N=8$ \tips{pe} for our final Spartus implementation. 
For Edge-Spartus, we set $M=4$ and $N=1$ to have a similar off-chip \tip{dram} bandwidth as the EdgeDRNN accelerator (EdgeDRNN requires 64 bits per clock cycle while Edge-Spartus requires 72 bits per clock cycle) for a relatively fair comparison in Sections~\ref{sec:compare_spartus} and~\ref{sec:compare_edge}.
\subsection{Model Accuracy}
\label{sec:model_acc}
We evaluated the final accuracy of our \tip{am} using our proposed training and optimization methods on the TIMIT and Librispeech test sets with $M=64$ and different delta thresholds.
\subsubsection{TIMIT}
To demonstrate the effectiveness of \tip{cbtd}, we trained the \tips{am} with various numbers of layers and layer sizes on the TIMIT dataset, as shown in Table~\ref{tab:timit}. 
The largest network has two unidirectional \tip{lstm} layers with 1024 units per layer, which is the same as the networks used in \tip{ese}~\cite{han2017ese} and other related previous works, including C-\tip{lstm}~\cite{Wang2018}, E-\tip{rnn}~\cite{Li2019}, \tip{bbs}~\cite{Cao2019} and E-\tip{lstm}~\cite{Wang2019}. 
We observe that networks pretrained with our \tip{cbtd} method result in better \tip{per} on the test set than the corresponding FP32 baselines when the spatial (weight) sparsity is roughly between 80\% and 94\%. 
Networks are likely to have worse \tip{per} when the spatial sparsity is $\ge 97$\%.
These results on the TIMIT test set are consistent with those shown in Fig~\ref{Fig:Explore_Array} on the development set. 
In short, our \tip{cbtd} method achieves 16$\times$ lossless compression of the various sizes of \tips{am}. The previous works either could not achieve the same level of model compression rate or had worse accuracy loss. 
By further adding temporal sparsity on top of spatial sparsity, our spatio-temporal \tip{deltalstm} network achieves 170$\times$ saving of arithmetic operations and memory access of the \tip{mxv} in the \tip{lstm} layers, which is significantly better than previous works.
\begin{figure}[!t]
	\centering
	\includegraphics[width=0.8\linewidth]{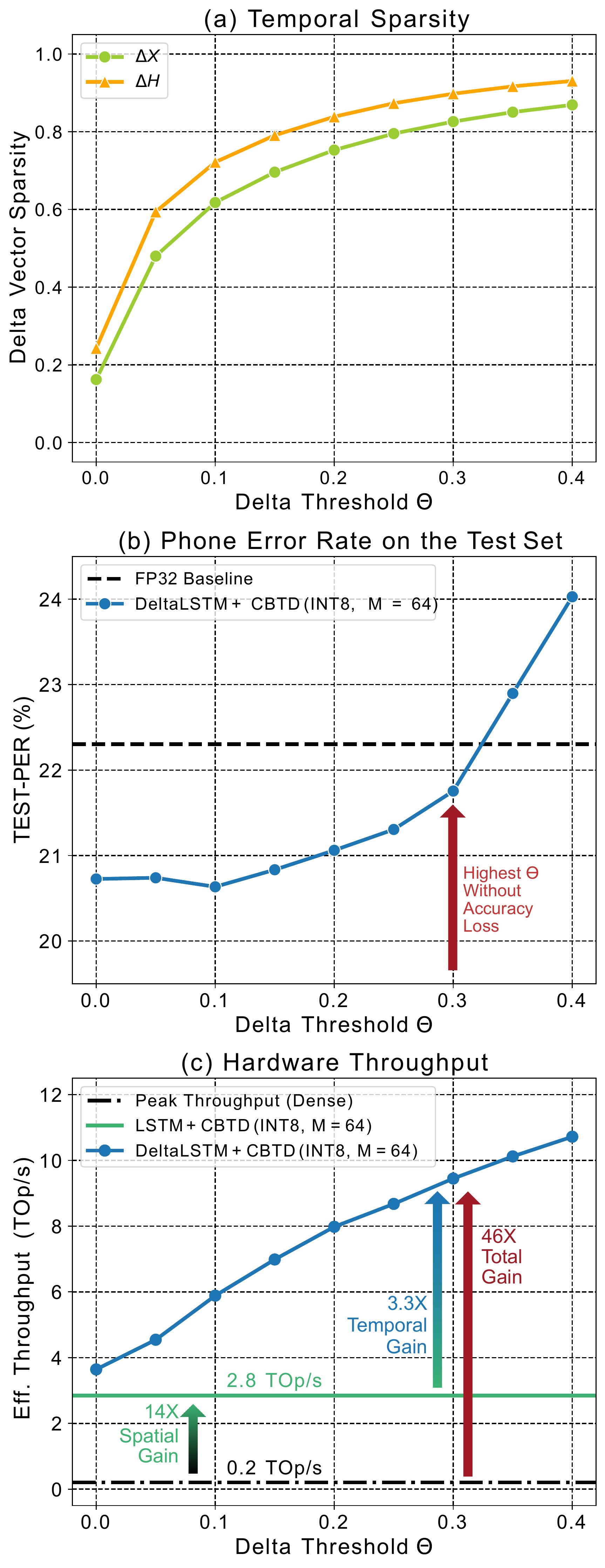}
	\caption{(a) shows the temporal sparsity of the \tip{deltalstm}-2L-1024H-UNI-CBTD network. $\Delta x$ and $\Delta h$ are respectively the delta input and delta hidden vectors;(b) shows the \tip{per} of the \tip{deltalstm} network evaluated on the core test set of TIMIT;  (c) shows the batch-1 throughput of the Spartus accelerator versus delta thresholds. The theoretical peak throughput is computed using \eqref{eq:peak_throughput}.}
	\label{Fig:delta}
\end{figure}
Fig.~\ref{Fig:delta}(a) shows the impact of different delta threshold values on the temporal sparsity evaluated on the TIMIT core test set, and the values are the average temporal sparsity over the 2 \tip{deltalstm} layers.
Overall, the sparsity of delta input state vector $\Delta x$ is lower than delta hidden state vector $\Delta h$. 
This is due to the small temporal sparsity in the input layer of the whole network to allow enough input stimuli to be fed into the network. 
The delta hidden states can be up to 90\% sparse when the threshold is larger than 0.3.
Fig.~\ref{Fig:delta}(b) shows the \tip{per} on the core test set (\tip{test}-\tip{per}) versus the delta threshold.
The FP32 baseline on the test set was 22.30\%. 
Similar to \tip{dev}-\tip{per} results, the \tip{deltalstm} network achieves better \tip{test}-\tip{per} when the delta threshold $\Theta$ is smaller than 0.3. 
With small delta thresholds, the \tip{deltalstm} shows a slight regularization effect that helped to achieve the best \tip{test}-\tip{per} at 20.63\%.
With $\Theta=0.3$, the \tip{test}-\tip{per} is 21.75\% which corresponds to $0.55\%$ accuracy improvement compared to the FP32 baseline.
\subsubsection{Librispeech} \label{sussubsec:librispeech}
Beyond the simpler TIMIT dataset, we also evaluate the effectiveness of cost saving and accuracy impact of our methods on the large-scale corpus, Librispeech, which has not been evaluated in previous \tip{fpga} \tip{rnn} accelerators. 
Results are shown in Table~\ref{tab:librispeech}. 
We first applied \tip{cbtd} on the bidirectional \tip{lstm} used in the PyTorch-Kaldi baseline~\cite{Ravanelli2019} and obtained 8$\times$ arithmetic operations saving with 0.2\% better \tip{wer} than the baseline. 
16$\times$ saving is achievable with 0.7\% accuracy loss. 
The bidirectional \tip{lstm} networks achieve the highest accuracy but are not usable in real-time speech recognition systems because they must receive the whole sentence to recognize a single frame. 
Thus, we are more interested in evaluating our methods on unidirectional \tip{lstm} networks. 
To our knowledge, \tip{hcgs}~\cite{Kadetotad2020} is the only previous structured pruning method evaluated on Librispeech. 
However, the sparse weight pattern of \tip{hcgs} is predefined at the network initialization stage and maintained throughout training with only changeable weight magnitudes, greatly limiting its ability to find subnets with good accuracy, as discussed in \textit{The Lottery Ticket Hypothesis}~\cite{frankle2018,Malach2020}. 
\tip{cbtd} allows both removal and addition of connections during training with only constraints on the distribution of nonzero values across predefined submatrices, which could lead to higher accuracy with the same level of model compression. 
We applied \tip{cbtd} on a unidirectional \tip{lstm} network of the same size as reported in ~\cite{Kadetotad2020}. 
With the same 16$\times$ saving in arithmetic operations, \tip{cbtd} has only 0.3\% higher \tip{wer} while \tip{hcgs} caused 10$\times$ higher accuracy loss. 
Finally, we applied our methods to the \tip{am} model used for hardware evaluation. 
We found that the combination of INT8 quantization, \tip{cbtd}, \tip{deltalstm} achieves 64$\times$ model compression and 81.8$\times$ arithmetic operations saving compared to the corresponding FP32 baseline. 
Another important observation is that networks pruned by \tip{cbtd} during training with conservative pruning rates (less or equal to 94\% for TIMIT and 90\% for Librispeech) all led to better accuracy, proving that \tip{cbtd} well preserves the regularization effect of pruning, as suggested by related works~\cite{han2016dsd,frankle2018,Malach2020}.

\subsection{Hardware Performance: Throughput \& Latency} \label{sec:throughput-latency}
\begin{table}[!t]
\centering
\caption{Performance summary of Spartus and Edge-Spartus with progressive levels of optimizations (Sec.~\ref{sec:throughput-latency})}
\resizebox{0.48\textwidth}{!}{ 
\begin{tabular}{|l|c|cc|cc|}
\hline
\multicolumn{1}{|c|}{\multirow{2}{*}{\textbf{}}} & \multirow{2}{*}{\textbf{Settings}} & \multicolumn{2}{c|}{\textbf{\begin{tabular}[c]{@{}c@{}}Batch-1 Eff. \\ Throughput\\ (GOp/s)\end{tabular}}} & \multicolumn{2}{c|}{\textbf{\begin{tabular}[c]{@{}c@{}}Latency\\ ($\mu$s)\end{tabular}}} \\ \cline{3-6} 
\multicolumn{1}{|c|}{}                           &                                    & \multicolumn{1}{c|}{\textbf{Spartus}}                               & \textbf{Edge}                           & \multicolumn{1}{c|}{\textbf{Spartus}}                    & \textbf{Edge}                    \\ \hline
\textbf{No Opt.}                                 & -                                  & \multicolumn{1}{c|}{\textless{}204.8}                           & \textless{}1.0                           & \multicolumn{1}{c|}{\textgreater{}46.0}              & \textgreater{}9420.8              \\ \hline
\textbf{+\tip{cbtd}}                                   & $\gamma=0.94$                      & \multicolumn{1}{c|}{2845.4}                                     & 17.7                                     & \multicolumn{1}{c|}{3.3}                             & 530.2                             \\ \hline
\multirow{2}{*}{\textbf{+\tip{deltalstm}}}             & $\Theta=0.1$                       & \multicolumn{1}{c|}{5885.0}                                     & 34.8                                     & \multicolumn{1}{c|}{1.6}                             & 270.6                             \\ \cline{2-6} 
                                                 & $\Theta=0.3$                       & \multicolumn{1}{c|}{9447.8}                                     & 77.3                                     & \multicolumn{1}{c|}{1.0}                             & 121.7                             \\ \hline
\end{tabular}
}
\label{tab:summary}
\end{table}
Since we focus on achieving low latency \tip{lstm} inference, in this work, we evaluate the hardware performance with a batch size of 1. The test network is the top layer of our biggest \tip{am} having 1024 hidden units, which is the same as previous state-of-the-art \tip{fpga} \tip{rnn} accelerators~\cite{han2017ese,Wang2018,Li2019,Cao2019,Wang2019}.

Table~\ref{tab:summary} summarized the performance of Spartus and Edge-Spartus with progressive levels of optimizations.
With $M\times N=512$ \tip{mac} units, the theoretical baseline performance of Spartus is only 0.2 TOp/s, calculated using \eqref{eq:peak_throughput}.
With the highest level of optimization, Spartus can finish the inference of the big \tip{deltalstm} layer within 1 $\mu$s, corresponding to 9.4\,TOp/s effective throughput and 46$\times$ speedup versus the baseline. 
Fig.~\ref{Fig:delta}(c) further shows how much the spatial and temporal sparsity contributes to the total speedup. 
By applying only \tip{cbtd} to the \tip{lstm} network to induce 94\% spatio sparsity (16$\times$ pruning rate), the Spartus achieves 2.8 TOp/s batch-1 throughput, corresponding to a 14X spatial gain the baseline.
After retraining the sparse \tip{lstm} layers as \tip{deltalstm}, further speedup was achieved. 
With a zero delta threshold, the accelerator achieves 3.6 TOp/s batch-1 throughput. 
By increasing the delta threshold to 0.3, which is the highest value with better \tip{test}-\tip{per} over FP32 baseline on TIMIT, Spartus achieves 9.4 TOp/s batch-1 throughput, which is another 3.3$\times$ temporal gain on top of spatial gain.
Overall, by combining the 14 $\times$ spatial gain and $3.3\times$ temporal gain, Spartus achieves 46$\times$ speedup by exploiting spatio-temporal sparsity.

Table~\ref{tab:summary} also reports the performance of Edge-Spartus, which was implemented on the \tip{fpga} of the smallest Zynq \tip{soc}, which has only 0.2\,MB on-chip memory and cannot buffer the networks on-chip even after compression. 
By relying solely on off-chip DDR3L memory bandwidth to fetch the network weights, Edge-Spartus achieves 121.7\,$\mu$s latency, which is still orders of magnitude faster than the normal 10\,ms frame shift used in real-time speech recognition systems~\cite{Amodei2015,Gao2019iscas}

\section{Discussion}
\subsection{Comparing Spartus with Previous Works on TIMIT}
\label{sec:compare_spartus}
\begin{table*}[!t]
  \caption{Comparison of Spartus with prior state-of-the-art \tipshort{rnn} accelerators on \tip{fpga} (Batch Size = 1)}
  \label{tab:compare}
  \centering
\begin{threeparttable}
\resizebox{0.98\textwidth}{!}{ 
\begin{tabular}{|l|c|c|cccc|c|}
\hline
                                                                                           & \textbf{\tip{ese}~\cite{han2017ese}} & \textbf{DeltaRNN~\cite{GaoDeltaRNN2018}} & \multicolumn{1}{c|}{\textbf{C-\tip{lstm}~\cite{Wang2018}}} & \multicolumn{1}{c|}{\textbf{E-\tipshort{rnn}~\cite{Li2019}}} & \multicolumn{1}{c|}{\textbf{BBS~\cite{Cao2019}}} & \textbf{E-\tip{lstm}~\cite{Wang2019}} & \textbf{Spartus (Ours)} \\ \hline
\textbf{\#Parameters (M)}                                                                  & 3.25         & 0.20              & \multicolumn{1}{c|}{3.25}            & \multicolumn{1}{c|}{3.25}           & \multicolumn{1}{c|}{3.25}         & 4.82            & \textbf{4.70}               \\ \hline
\textbf{Compressed \#Parameters (M)}                                                       & 0.36         & -                 & \multicolumn{2}{c|}{0.20}                                                  & \multicolumn{1}{c|}{0.41}         & 0.60            & \textbf{0.29}               \\ \hline
\textbf{\begin{tabular}[c]{@{}l@{}}Bit Precision\\ (Activation/Weight/Index)\end{tabular}} & INT16/12/4   & INT16/16/0        & \multicolumn{1}{c|}{INT6/16/0}       & \multicolumn{1}{c|}{INT16/16/0}     & \multicolumn{1}{c|}{INT16/16/4}   & INT8/8/4        & \textbf{INT16/8/8}          \\ \hline
\textbf{Weight Sparsity (\%)}                                                              & 88.78        & 0                 & \multicolumn{2}{c|}{93.75}                                                 & \multicolumn{2}{c|}{87.5}                           & \textbf{93.75}              \\ \hline
\textbf{Act./Temp. Sparsity (\%)}                                                          & 0            & 92.5              & \multicolumn{3}{c|}{0}                                                                                         & 58              & \textbf{82.56}              \\ \hline
\textbf{PER on TIMIT (\%)}                                                                 & 20.7         & -                 & \multicolumn{1}{c|}{24.6}            & \multicolumn{1}{c|}{20.3}           & \multicolumn{1}{c|}{23.6}         & 23.2            & \textbf{21.8$\pm$0.3}       \\ \hline
\textbf{Language Model?}                                                              & Yes          & -                 & \multicolumn{3}{c|}{Unreported}                                                                                & No              & \textbf{No}                 \\ \hline \hline
\textbf{FPGA Platform}                                                                     & XCKU060      & XC7Z100           & \multicolumn{1}{c|}{7V3}             & \multicolumn{1}{c|}{XC7VX690T}      & \multicolumn{1}{c|}{GX1150}       & SX660           & \textbf{XC7Z100}            \\ \hline
\textbf{DSP (\%)}                                                                          & 54.5         & 38.0              & \multicolumn{1}{c|}{74.3}            & \multicolumn{1}{c|}{79.6}           & \multicolumn{1}{c|}{100}          & 1.4             & \textbf{25.7}               \\ \hline
\textbf{BRAM/M20K (\%)}                                                                    & 87.7         & 60.6              & \multicolumn{1}{c|}{65.7}            & \multicolumn{1}{c|}{65.2}           & \multicolumn{1}{c|}{92}           & 32.1            & \textbf{33.0}               \\ \hline
\textbf{LUT/ALM (\%)}                                                                      & 88.6         & 94.2              & \multicolumn{1}{c|}{58.7}            & \multicolumn{1}{c|}{59.4}           & \multicolumn{1}{c|}{68}           & 87.8            & \textbf{49.2}               \\ \hline
\textbf{FF (\%)}                                                                           & 68.3         & 21.5              & \multicolumn{1}{c|}{46.5}            & \multicolumn{1}{c|}{55.3}           & \multicolumn{1}{c|}{N/A}          & 15.6            & \textbf{19.5}               \\ \hline \hline
\textbf{Frequency (MHz)}                                                                   & 200          & 125               & \multicolumn{4}{c|}{200}                                                                                                         & \textbf{200}                \\ \hline
\textbf{\#MACs}                                                                            & 32           & 768               & \multicolumn{2}{c|}{128}                                                   & \multicolumn{1}{c|}{4096}         & 128             & \textbf{512}                \\ \hline
\textbf{Peak Throughput (GOp/s)}                                                           & 12.8         & 192               & \multicolumn{2}{c|}{51.2}                                                  & \multicolumn{1}{c|}{1638.4}       & 51.2            & \textbf{204.8}              \\ \hline
\textbf{Effective Throughput (GOp/s)}                                                      & 78.6         & 1198.0            & \multicolumn{1}{c|}{714.3}           & \multicolumn{1}{c|}{783.1}          & \multicolumn{1}{c|}{2432.8}       & 403.3           & \textbf{9447.8}             \\ \hline
\textbf{Speedup}                                                                           & 6.1$\times$  & 6.2$\times$       & \multicolumn{1}{c|}{14.0$\times$}    & \multicolumn{1}{c|}{15.3$\times$}   & \multicolumn{1}{c|}{1.5$\times$}  & 7.9$\times$     & \textbf{46.1$\times$}       \\ \hline
\textbf{Latency (us)}                                                                      & 82.7         & -                 & \multicolumn{1}{c|}{9.1}             & \multicolumn{1}{c|}{8.3}            & \multicolumn{1}{c|}{2.4}          & 23.9            & \textbf{1.0}                \\ \hline
\textbf{Frame per Second (kFPS)}                                                           & 12           & -                 & \multicolumn{1}{c|}{110}             & \multicolumn{1}{c|}{120}            & \multicolumn{1}{c|}{417}          & 42              & \textbf{1,001}              \\ \hline
\textbf{Wall Power (W)}                                                                    & 41.0         & 7.3               & \multicolumn{1}{c|}{23.0}            & \multicolumn{1}{c|}{25.0}           & \multicolumn{1}{c|}{19.1}         & 15.9            & \textbf{8.4}                \\ \hline
\textbf{Wall Power Efficiency (\tnote{1}~GOp/s/W)}                                                   & 1.9          & 164.1             & \multicolumn{1}{c|}{31.1}            & \multicolumn{1}{c|}{31.3}           & \multicolumn{1}{c|}{127.4}        & 25.4            & \textbf{1124.7}             \\ \hline
\end{tabular}
}
\begin{tablenotes}\footnotesize
\item[1] GOp/s/W is equivalent to GOp/J.

\end{tablenotes}

\end{threeparttable}
\end{table*}
\begin{table}[!t]
  \caption{Comparison of Edge-Spartus with EdgeDRNN (Batch Size = 1)}
  \label{tab:compare_small}
  \centering
\resizebox{0.48\textwidth}{!}{ 
\begin{tabular}{|l|cc|}
\hline
                                                                                           & \multicolumn{1}{c|}{\textbf{EdgeDRNN~\cite{edgedrnn}}} & \textbf{Edge-Spartus (Ours)} \\ \hline
\textbf{\#Parameters (M)}                                                                  & \multicolumn{1}{c|}{5.4}               & \textbf{4.7}                  \\ \hline
\textbf{Compressed \#Parameters (M)}                                                       & \multicolumn{1}{c|}{-}                 & \textbf{0.29}                 \\ \hline
\textbf{\begin{tabular}[c]{@{}l@{}}Bit Precision\\ (Activation/Weight/Index)\end{tabular}} & \multicolumn{1}{c|}{INT16/8/0}         & \textbf{INT16/8/10}           \\ \hline
\textbf{Weight Sparsity (\%)}                                                              & \multicolumn{1}{c|}{0}                 & \textbf{93.75}                \\ \hline
\textbf{Act./Temp. Sparsity (\%)}                                                          & \multicolumn{1}{c|}{90.01}             & \textbf{82.56}                \\ \hline \hline
\textbf{FPGA Platform}                                                                     & \multicolumn{2}{c|}{\textbf{XC7Z007S}}                                 \\ \hline
\textbf{DSP (\%)}                                                                          & \multicolumn{1}{c|}{13.6}              & \textbf{7.6}                  \\ \hline
\textbf{BRAM (\%)}                                                                         & \multicolumn{1}{c|}{66.0}              & \textbf{76.0}                 \\ \hline
\textbf{LUT(\%)}                                                                           & \multicolumn{1}{c|}{65.2}          
    & \textbf{78.5}                 \\ \hline
\textbf{FF (\%)}                                                                           & \multicolumn{1}{c|}{34.1}              & \textbf{41.5}                 \\ \hline \hline
\textbf{Frequency (MHz)}                                                                   & \multicolumn{2}{c|}{\textbf{125}}                                      \\ \hline
\textbf{\#MACs}                                                                            & \multicolumn{1}{c|}{8}                 & \textbf{4}                    \\ \hline
\textbf{Latency (us)}                                                                      & \multicolumn{1}{c|}{536}               & \textbf{121.7}                \\ \hline
\textbf{Frame per Second (kFPS)}                                                           & \multicolumn{1}{c|}{1.9}               & \textbf{8.2}                  \\ \hline
\textbf{Peak Throughput (GOp/s)}                                                           & \multicolumn{1}{c|}{2.0}               & \textbf{1.0}                  \\ \hline
\textbf{Effective Throughput (GOp/s)}                                                      & \multicolumn{1}{c|}{20.2}              & \textbf{77.3}                 \\ \hline
\textbf{Speedup}                                                                           & \multicolumn{1}{c|}{10.1$\times$}      & \textbf{77.3$\times$}         \\ \hline
\textbf{Accelerator Core Power ($\mu$W)}                                                   & \multicolumn{1}{c|}{66}                & \textbf{69}                   \\ \hline
\textbf{Core Power Efficiency (GOp/s/W)}                                                   & \multicolumn{1}{c|}{306}               & \textbf{1120}                 \\ \hline
\textbf{FPGA On-Chip Power (W)}                                                            & \multicolumn{2}{c|}{\textbf{1.4}}                                      \\ \hline
\textbf{On-Chip Power Efficiency (GOp/s/W)}                                                & \multicolumn{1}{c|}{14.2}              & \textbf{53.3}                 \\ \hline
\textbf{Wall Power (W)}                                                                    & \multicolumn{2}{c|}{\textbf{2.3}}                                      \\ \hline
\textbf{Wall Power Efficiency (GOp/s/W)}                                                   & \multicolumn{1}{c|}{8.8}               & \textbf{33.6}                 \\ \hline
\end{tabular}
}
\end{table}
\tip{fpga}
\tip{ese} was the first \tipshort{rnn} accelerator that adopted weight pruning to speedup \tip{lstm} inference. 
However, \tip{ese} was designed for throughput-oriented inference using large batch sizes. 
Performance of \tip{ese} was tested with a batch size of 32.
DeltaRNN exploited temporal sparsity to accelerate \tipshort{gru} RNNs, but the test network was small compared to other works.
C-\tip{lstm} and E-\tipshort{rnn} used a structured weight matrix with an FFT-based computing mechanism to reduce operations during inference. 
Table~\ref{tab:compare} provides the performance of C-\tip{lstm} and E-\tipshort{rnn} with the 16$\times$ compression ratio reported in the original papers.
E-\tip{lstm} and BBS adopted structured pruning methods to achieve finer-grained workload balance compared to \tip{ese}, and their performance was evaluated with batch sizes of 8 and 1, respectively. 
BBS achieved the best batch-1 throughput and latency among all previous \tipshort{rnn} accelerators.
Since we focus on achieving low-latency \tip{lstm} inference, for a fair comparison, we computed the performance of \tip{ese} and E-\tip{lstm} with a batch size of 1 in Table~\ref{tab:compare}. 
Power numbers of Spartus and other accelerators are measured using the wall-plug power of the \tip{fpga} board.

We include 'Latency' and 'Frame per Second (FPS)' to represent the performance of accelerators on their own test networks. 
Spartus achieves the highest FPS and lowest inference latency among all platforms.
However, the test network sizes are different across platforms. 
Spartus and E-\tip{lstm} used test networks with normal \tip{lstm} units, while \tip{ese}, C-\tip{lstm}, E-\tipshort{rnn}, and BBS used a Google \tip{lstm} network with the same number of neurons but because their neuron model has peephole connections and projection layers that shrink the dimension of both recurrent and output connections~\cite{sak2014long}, leading to an overall decrease in the number of network parameters.
Thus, we include 'Effective Throughput' and 'Power Efficiency' metrics which take into the number of total operations for computing these networks in the comparison across these accelerators. Compared to DeltaRNN, which also exploited temporal sparsity, Spartus achieves 8$\times$ higher effective throughput and 7$\times$ higher power efficiency. 
Compared to C-\tip{lstm} and E-\tipshort{rnn} which achieved the same 16$\times$ weight compression ratio, Spartus achieves around 10$\times$ higher FPS and 8$\times$ lower latency. 
By exploiting spatio-temporal sparsity, Spartus achieves 4$\times$ higher batch-1 effective throughput than BBS, which was the state-of-the-art accelerator with the highest batch-1 effective throughput.

Table~\ref{tab:compare} also includes the resource utilization and names of the \tip{fpga} chips. 
Table~\ref{tab:fpga} compares \tips{fpga}. 
A larger \tip{fpga} can buffer larger networks on-chip to provide sufficient memory bandwidth to more physical \tip{mac} units.
New process technologies can help the same accelerator architecture to achieve higher clock frequencies after implementation.
It can be observed that Spartus achieves higher performance than previous accelerators even on a smaller \tip{fpga} with the earlier 28nm process, except for DeltaRNN that used the same XC7Z100 \tip{fpga}.
\subsection{Comparing Edge-Spartus to EdgeDRNN and \tip{ese}}
\label{sec:compare_edge}
Most previous accelerators achieve their high batch-1 throughput by storing the \tipshort{rnn} weights completely on-chip, which is not practical in edge applications using resource-constrained hardware platforms. 
EdgeDRNN~\cite{edgedrnn} and \tip{ese}~\cite{han2017ese} are the only two previous \tip{fpga} \tip{rnn} accelerators that reported real performance numbers with the off-chip memory bandwidth bottleneck.
Thanks to the scalable architecture of Spartus, we could fit Edge-Spartus with 4 \tips{pe} on the tiny \tip{fpga} in the Zynq XC7007S \tip{soc}. 
Edge-Spartus uses INT8 weights with a 10-bit local index for each weight; thus, the off-chip memory interface bit-width is 72-bits; while the previous EdgeDRNN with 8 \tips{pe} used INT8 weights and requires a 64-bit off-chip memory interface.
With the same DDR3L off-chip memory as EdgeDRNN on the MiniZed board, Edge-Spartus achieves 77.3\,GOp/s batch-1 effective throughput, which is around 4$\times$ higher than EdgeDRNN and close to that of \tip{ese}, which uses 8$\times$ more \tip{mac} units on an \tip{fpga} that costs 70$\times$ more.
\subsection{DRAM Power Efficiency}
\begin{figure}[!t]
	\centering
	\includegraphics[width=0.9\linewidth]{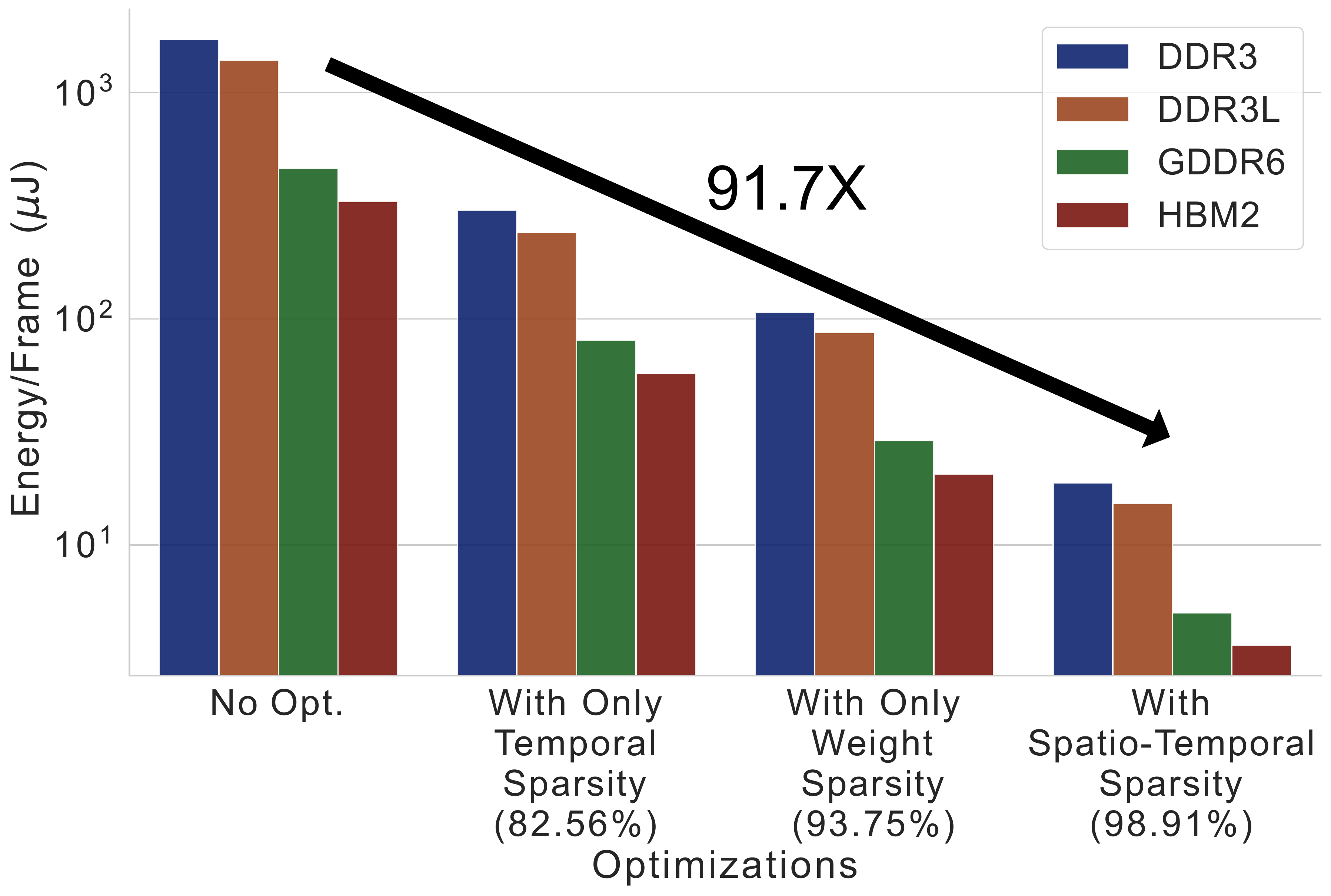}
	\caption{Estimated Off-chip \tip{dram} access energy of Edge-Spartus per inference (Frame).}
	\label{Fig:dram}
\end{figure}

On-chip memory fetch consumes over 10$\times$ higher energy than arithmetic operations with the same numbers of bits and can be over 1000$\times$ when using off-chip \tip{dram}~\cite{Horowitz2014,Jouppi2021}.
For a memory-bounded algorithm such as \tip{lstm}, the key to enhancing power efficiency is to reduce the amount of memory access per inference. Using Edge-Spartus, we analyzed the potential reduction of energy consumption per frame using various types of \tip{dram} (Table~\ref{tab:dram}) and showed the results in Fig.~\ref{Fig:dram}. By exploiting spatio-temporal sparsity, the \tip{dram} access energy can be reduced by 91.7$\times$. 
However, this reduction of energy does not completely translate to the power efficiency of the whole system, of which other peripheral modules can also be an important source of power consumption. 
For example, our XC7Z100 \tip{som} lacks a usable \tip{ps} sleep mode and thus consumes 1.8\,W even at idle. Thus, although the power efficiency obtained using wall power reflect the real-world performance of different accelerators, they are biased by the boards they are implemented on. 
Our comparison between EdgeDRNN and Edge-Spartus showcases how much improvement of power efficiency can be achieved on the same platform by exploiting spatio-temporal sparsity, and Edge-Spartus achieves roughly 4$\times$ higher core and system power efficiency than EdgeDRNN.

\begin{table}[!t]
\centering
  \caption{Off-chip DRAM access energy.}
  \label{tab:dram}
\begin{threeparttable}
\begin{tabular}{|l|l|l|l|l|}
\hline
\textbf{DRAM Type}              & \textbf{DDR3}             & \textbf{DDR3L}            & \textbf{GDDR6}           & \textbf{HBM2}            \\ \hline
\textbf{Access Energy/Bit (pJ)} & \multicolumn{1}{c|}{20.3~\cite{Horowitz2014}} & \multicolumn{1}{c|}{\tnote{1}~16.5} & \multicolumn{1}{c|}{5.5~\cite{gddr6}} & \multicolumn{1}{c|}{3.9~\cite{Connor2017}} \\ \hline
\end{tabular}
\begin{tablenotes}\footnotesize
\item[1]The DDR3L (1.35\,V) access energy is estimated by scaling down the number of DDR3 (1.5\,V) according to their supply voltages. 
\end{tablenotes}
\end{threeparttable}
\end{table}

\section{Conclusion}

We proposed Spartus, the first \tip{lstm} accelerator that exploits spatio-temporal sparsity to enable the lowest latency and the highest power efficiency in \tipshort{rnn} inference compared to previous work. The spatial weight sparsity was implemented using our newly proposed structured \tip{cbtd} pruning method.
\tip{cbtd} achieves a comparable compression rate as previous weight compression methods with negligible accuracy loss. 
The benefit of \tip{cbtd} over previous methods is the compatibility with the \tip{dn} algorithm.
Temporal sparsity is achieved through the \tip{deltalstm} model by applying the delta network algorithm to the \tip{lstm} model.
The Spartus accelerator is implemented on a Xilinx Zynq 7100 \tip{fpga} running at 200\;MHz. 
Evaluated on the TIMIT dataset, Spartus achieves 9.4 TOp/s effective batch-1 throughput and 1.1 TOp/J power efficiency, which is respectively 4$\times$ and 7$\times$ higher than the previous state-of-the-art. 
Compared to the theoretical peak hardware performance, which runs a dense \tip{lstm} layer with 1024 neurons in 46\,\textmu s, Spartus runs the same network in 1\,\textmu s by jointly exploiting structured spatial sparsity (14$\times$ speedup) and temporal sparsity (3.3$\times$ speedup) to achieve 46$\times$ speedup in total.
For light-weight and low-cost edge applications, Edge-Spartus provides over 75~GOp/s effective throughput for arbitrary-sized \tip{lstm}-\tips{rnn} on a \$55 USB-powered MiniZed board, which will be useful for edge signal processing and mobile robots. 
Compared to \tip{asic}-based \tip{rnn} accelerators~\cite{Kadetotad2020} which mainly focus on low power processing, Spartus achieves significantly higher effective throughput but also massively more power consumption due to its implementation in \tips{fpga}.
A future \tip{asic} implementation of Spartus could achieve even higher power efficiency and throughput than the existing \tip{fpga}-based systems.

\section*{Acknowledgment}
\small{We thank X. Chen and other Sensors Group members for discussion on general design concepts. We thank A. Rios-Navarro, R. Morales, and A. Linares-Barranco from the University of Seville for creating the baseboard for our FPGAs. }


%


\ifCLASSOPTIONcaptionsoff
  \newpage
\fi


\renewcommand*{\bibfont}{\footnotesize}

\printbibliography
\vspace{-1.5cm}
\begin{IEEEbiography}[{\includegraphics[width=1in,height=1.25in,clip,keepaspectratio]{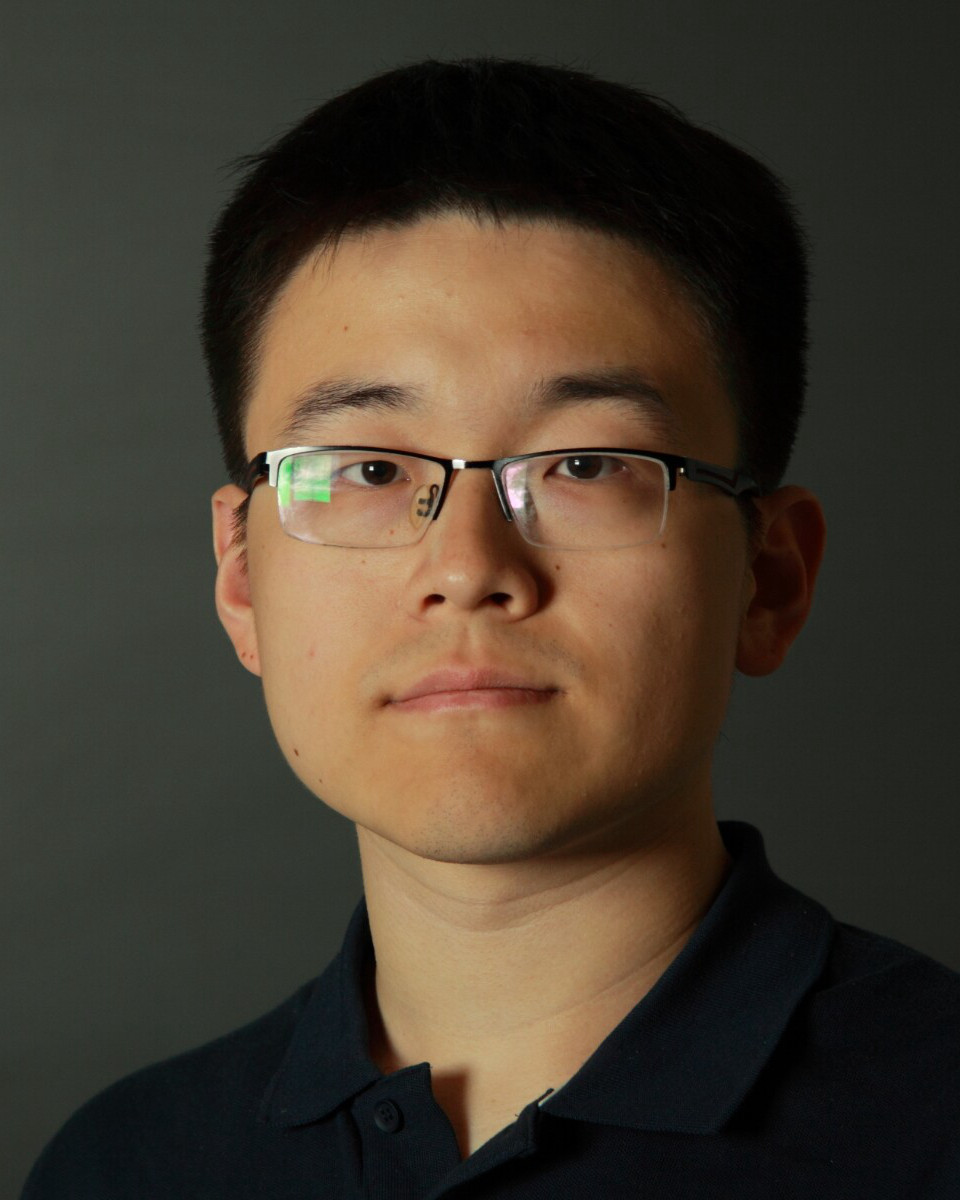}}]{Chang Gao}
(Member, IEEE) received the BEng degree in Electronics from University of Liverpool, Liverpool,
UK and Xi'an Jiaotong-Liverpool University, Suzhou, China, and the MSc degree in analog and digital integrated circuit design from Imperial College London, London, UK. He was awarded his Doctoral degree at the Institute of Neuroinformatics, University of Zurich and ETH Zurich, Zurich, Switzerland in Dec. 2021 and is joining TU Delft as an Assistant Professor in 2022.
His current research interests include computer architectures for deep learning with emphasis on recurrent neural networks.
\end{IEEEbiography}
\vspace{-1.5cm}
\begin{IEEEbiography}[{\includegraphics[width=1in,height=1.25in,clip,keepaspectratio]{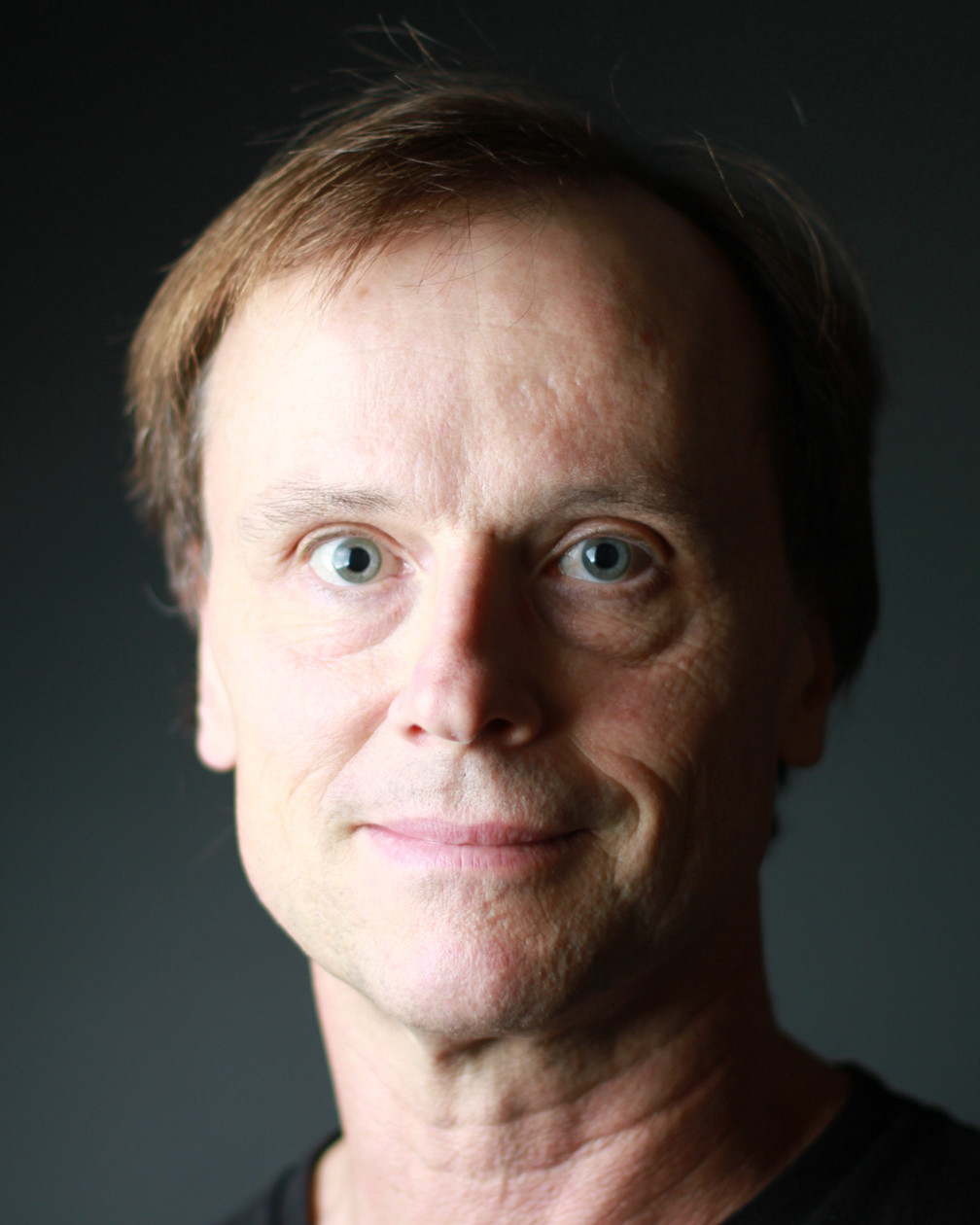}}]{Tobi Delbruck}
(Fellow, IEEE) received the
B.Sc. degree in physics from the University of
California at San Diego, San Diego, CA, USA,
in 1986, and the Ph.D. degree from the California
Institute of Technology, Pasadena, CA, USA,
in 1993.
Since 1998, he has been with the Institute of
Neuroinformatics, University of Zurich and ETH
Zurich, Zürich, Switzerland, where he is a
Professor of physics and electrical engineering. His
group along with S.-C. Liu focuses on neuromorphic
sensory processing, efficient hardware AI, and control for robotics.
\end{IEEEbiography}
\vspace{-1.5cm}
\begin{IEEEbiography}[{\includegraphics[width=1in,height=1.25in,clip,keepaspectratio]{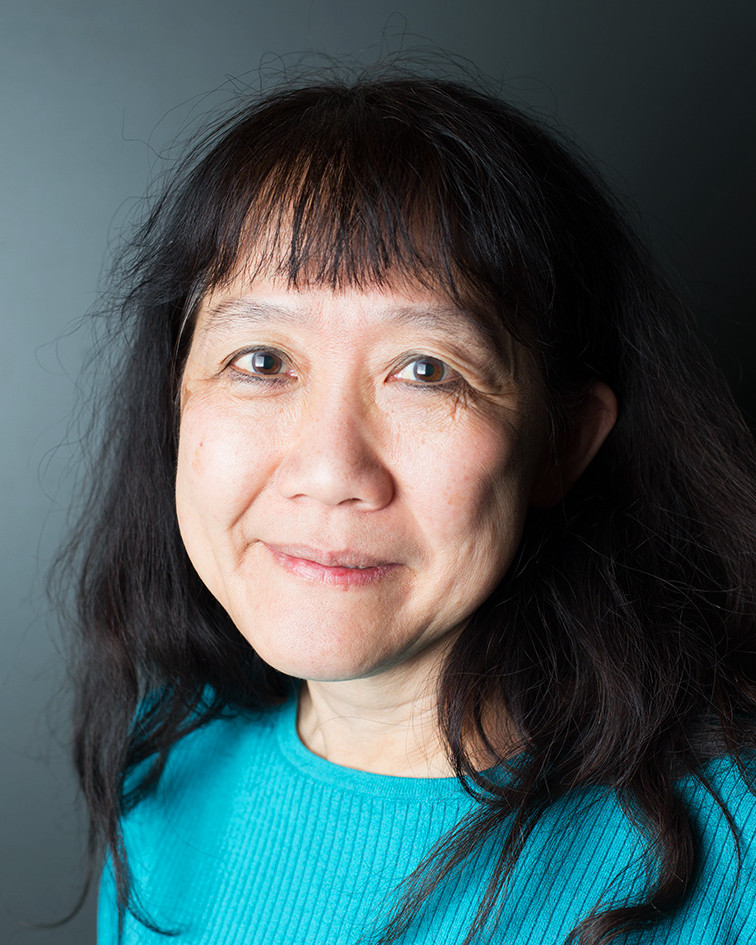}}]{Shih-Chii Liu}
(Fellow, IEEE) received the bachelor’s degree in electrical engineering from the
Massachusetts Institute of Technology, Cambridge,
MA, USA, and the Ph.D. degree in the computation
and neural systems program from the California
Institute of Technology, Pasadena, CA, USA,
in 1997.
She was with various companies, including Gould
American Microsystems, San Jose, CA, USA, LSI
Logic, Sherman Oaks, CA, USA, and Rockwell
International Research Labs, Thousand Oaks, CA,
USA. She is Professor and Group Leader at the Institute of Neuroinformatics,
University of Zurich and ETH Zurich, Zürich, Switzerland.
\end{IEEEbiography}




\end{document}